%% file: main.tex
\documentclass[10pt,conference]{IEEEtran}

\input{extrapackages}

\input{macros}

\title{An In-Memory Architecture for High-Performance Long-Read Pre-Alignment Filtering}

\author{
Taha Shahroodi\affilTUD{}\hspace{0.15in}%
Michael Miao\affilTUD{}\hspace{0.15in}%
Joel Lindegger\affilETH{}\hspace{0.15in}%
Stephan Wong\affilTUD{}\hspace{0.15in}%
Onur Mutlu\affilETH{}\hspace{0.15in}%
Said Hamdioui\affilTUD{}\vspace{2mm}\\%
\textit{\affilTUD{}TU Delft\hspace{0.30in}\affilETH{}ETH Z{\"u}rich}%
\\}

    \fancypagestyle{firststyle}
    {
        \setlength{\footskip}{40pt}
        \fancyfoot[C]{\thepage}
    }
\begin{document}
\maketitle

\input{Sections/00_abstract}

\input{Sections/01_introduction}

\input{Sections/02_background}

\input{Sections/03_motivation}

\input{Sections/05_algorithm}

\input{Sections/06_architecture_overview}

\input{Sections/07_tile_architecture}

\input{Sections/08_subarray_architecture}

\input{Sections/09_banksandbankgrouns_architecture}

\input{Sections/10_rank_architecture}

\input{Sections/11_algorithm_mapping}

\input{Sections/14_long_read_compatibility}

\input{Sections/13_differences_software_vs_hardware_Implementation}
\input{Sections/16_evaluation_methodology_setup}

\input{Sections/17_evaluations_and_results}

\input{Sections/19_relatedwork}

\input{Sections/20_conclusion}


\bibliographystyle{unsrt}
\bibliography{refs}


\end{document}

%% file: extrapackages.tex
\usepackage[square,sort,comma,numbers,compress]{natbib}

\usepackage{fancyhdr}
\usepackage{notoccite}

\usepackage{tikzsymbols}
\usepackage{array}
\usepackage{tablefootnote}
\usepackage{threeparttable}
\usepackage{upgreek}
\usetikzlibrary{shapes}
\usepackage{glossaries} 
\usepackage{xcolor}
\usepackage{ifthen}
\usepackage{makecell}
\usepackage{booktabs}
\usepackage[binary-units=true]{siunitx}[=v2]
\usepackage{hhline}
\usepackage{float}
\usepackage{algorithm}
\usepackage{algpseudocode}
\usepackage{datetime}
\usepackage{soul}
\usepackage{support-caption}

\usepackage[normalem]{ulem}
\usepackage{amsmath}
\usepackage{amssymb}
\usepackage{amsfonts}
\usepackage{cases}

\usepackage{marginnote} 

\usepackage{hyperref}
\usepackage{graphicx}
\usepackage{textcomp}
\usepackage{todonotes} 
\usepackage{textcomp}
\usepackage[T1]{fontenc}

\usepackage{balance}
\usepackage{flushend}

\usepackage{tcolorbox}

\usepackage{mathtools}
\usepackage[framemethod=tikz]{mdframed}
\usepackage{setspace}
\usepackage{blindtext}
\usepackage{imakeidx}

\usepackage{multicol}
\usepackage{multirow}
\title{Multicols Demo}
\usepackage{tabularx}

\usepackage{enumitem} 

%% file: macros.tex
\definecolor{lightyellow}{rgb}{0.980, 0.956, 0.623}
\sethlcolor{lightblue}
\definecolor{moegi}{rgb}{0.357, 0.537, 0.188}
\definecolor{amber}{rgb}{1.0, 0.49, 0.0}
\definecolor{awesome}{rgb}{1.0, 0.13, 0.32}
\definecolor{dollarbill}{rgb}{0.52,0.73,0.4}
\definecolor{moegi}{rgb}{0.357, 0.537, 0.188}
\definecolor{burgundy}{rgb}{0.5, 0.0, 0.13}
\definecolor{ballblue}{rgb}{0.13, 0.67, 0.8}
\definecolor{ups-truck}{rgb}{0.53, 0.28, 0.21}
\definecolor{airforceblue}{rgb}{0.36, 0.54, 0.66}
\definecolor{cadmiumgreen}{rgb}{0.0, 0.42, 0.24}
\definecolor{darkcyan}{rgb}{0.0, 0.55, 0.55}
\definecolor{caribbeangreen}{rgb}{0.0, 0.8, 0.6}
\definecolor{flamingopink}{rgb}{0.99, 0.56, 0.67}
\definecolor{jazzberryjam}{rgb}{0.65, 0.04, 0.37}
\definecolor{mediumpersianblue}{rgb}{0.0, 0.4, 0.65}
\definecolor{coolblack}{rgb}{0.0, 0.18, 0.39}
\definecolor{bleudefrance}{rgb}{0.19, 0.55, 0.91}
\definecolor{ao}{rgb}{0.0, 0.0, 1.0}
\definecolor{babyblueeyes}{rgb}{0.63, 0.79, 0.95}
\definecolor{darkwarmgray}{rgb}{0.2,0,0}
\definecolor{lightblue}{rgb}{0.980, 0.956, 0.623}

\usepackage{titlesec}
\titlespacing\section{0pt}{2pt plus 1pt minus 1pt}{3pt plus 1pt minus 2pt}
\titlespacing\subsection{0pt}{2pt plus 1pt minus 1pt}{3pt plus 1pt minus 2pt}
\titlespacing\subsubsection{0pt}{2pt plus 1pt minus 1pt}{3pt plus 1pt minus 2pt}

\makeatletter
\g@addto@macro{\normalsize}{%
\setlength{\abovedisplayskip}{1pt plus 1pt minus 1pt}
\setlength{\belowdisplayskip}{1pt plus 1pt minus 1pt}
\setlength{\abovedisplayshortskip}{0pt}
\setlength{\belowdisplayshortskip}{0pt}
\setlength{\intextsep}{1pt plus 1pt minus 1pt}
\setlength{\textfloatsep}{1pt plus 1pt minus 1pt}
\setlength{\skip\footins}{4pt plus 1pt minus 1pt}}
\makeatother
 
\newcommand{\boxbegin} {
	\begin{tcolorbox}[enhanced, frame hidden, colback=gray!50, breakable]
}

\newcommand{\boxend} {
	\end{tcolorbox}
}

\newcommand{\yboxbegin} {
	\begin{tcolorbox}[breakable, enhanced, frame hidden, colback=yellow!50]
}

\newcommand{\yboxend} {
	\end{tcolorbox}
}

\mdfdefinestyle{graybox}{
    splittopskip=0,%
    splitbottomskip=0,%
    frametitleaboveskip=0,
    frametitlebelowskip=0,
    skipabove=0,%
    skipbelow=0,%
    leftmargin=0,%
    rightmargin=0,%
    innertopmargin=0mm,%
    innerbottommargin=0mm,%
    roundcorner=1mm,%
    backgroundcolor=lightblue,
    hidealllines=true}
    
\mdfdefinestyle{graybox2}{
    splittopskip=0,%
    splitbottomskip=0,%
    frametitleaboveskip=0,
    frametitlebelowskip=0,
    skipabove=0,%
    skipbelow=0,%
    leftmargin=0,%
    rightmargin=0,%
    innertopmargin=2mm,%
    innerbottommargin=2mm,%
    roundcorner=2mm,%
    backgroundcolor=lightblue,
    hidealllines=true}

\newcommand{\bboxbegin}{
    \begin{mdframed}[style=graybox]
}

\newcommand{\bboxend}{
    \end{mdframed}
}

\newcommand{\yyboxbegin}{
    \begin{mdframed}[style=graybox2]
}

\newcommand{\yyboxend}{
    \end{mdframed}
}

\newcommand*\circled[1]{\tikz[baseline=(char.base)]{
            \node[shape=circle,draw,inner sep=0pt,fill=black, text=white] (char) {#1};}}

\newcommand*\Bcircled[1]{\tikz[baseline=(char.base)]{
            \node[shape=circle,draw,inner sep=0pt,fill=mediumpersianblue, text=white] (char) {#1};}}
\newcommand*\Ocircled[1]{\tikz[baseline=(char.base)]{
            \node[shape=circle,draw,inner sep=0pt,fill=orange, text=black] (char) {#1};}}
\newcommand*\Gcircled[1]{\tikz[baseline=(char.base)]{
            \node[shape=circle,draw,inner sep=0pt,fill=cadmiumgreen, text=white] (char) {#1};}}

\newcounter{obs}
\newcounter{take}
\newcounter{takeNN}
\newcommand\takeNN[1]{%
   \refstepcounter{takeNN}
   \begin{tcolorbox}[colback=white!25!white,colframe=black!65!white, arc=1pt, boxrule=0.5pt, left=2pt,right=2pt,top=0pt,bottom=0pt,  title=\textbf{{Takeaway}}]
   \emph{\takeawaychallengeobservationproblem{#1}}
   \end{tcolorbox}
}

\newcounter{obsbox}
\newcounter{chal}
\newcounter{chalbox}
\newcounter{problem}
\newcommand{\takeawaychallengeobservationproblem}[1]{\textcolor{black}{{#1}}}

\newif\ifcolored
\coloredtrue

\ifcolored

    \newcommand{\ignore}[1]{}

    \newcommand{\mstodo}[1]{\todo[size=\footnotesize,color=jazzberryjam]{\textbf{@M:} #1}}
    \newcommand{\mscomment}[1]{\textcolor{amber}{\textbf{[@Mike: {\it#1}]}}}
    \newcommand{\rccomment}[1]{\textcolor{magenta}{\textbf{[@Raphael: {\it#1}]}}}
    \newcommand{\promotorcomment}[1]{\textcolor{red}{\textbf{[@Promotor: {\it#1}]}}}
    \newcommand{\omcomment}[1]{\textcolor{red}{\textbf{[@OM: {\it#1}]}}}
    \newcommand\dtodo[1]{\noindent{\color{red} {\bf \fbox{\drWalley{}: }} {\it#1}}}
    \newcommand\fme[1]{\noindent{\color{red} {\bf \fbox{\drWalley{}: Fill me }} {\it#1}}}
    
\else

    \newcommand{\takeawaychallengeobservationproblem}[1]{{#1}}

    \newcommand{\ignore}[1]{}

    \newcommand{\mstodo}[1]{}
    \newcommand{\mscomment}[1]{}
    \newcommand{\rccomment}[1]{}
    \newcommand{\promotorcomment}[1]{}
    \newcommand{\omcomment}[1]{}
    \newcommand\dtodo[1]{}
    \newcommand\fme[1]{}
  
\fi

\newcommand\fig[1]{Fig.~{#1}\xspace}

\newcommand\sect[1]{Section~{#1}\xspace}

\newcommand\alg[1]{Algorithm~{#1}\xspace}

\newcommand\tab[1]{Table~{#1}\xspace}

\newcommand{\cpp}{\mbox{C++}\xspace}

\newcommand{\pim}{PIM\xspace}
\newcommand{\pimlong}{Processing-In-Memory\xspace}
\newcommand{\cim}{CIM\xspace}
\newcommand{\cimlong}{Computation-In-Memory\xspace}

\newcommand{\sram}{SRAM\xspace}

\newcommand{\dram}{DRAM\xspace}
\newcommand{\dramlong}{dynamic random-access memory\xspace}

\newcommand{\sotalong}{State-of-the-Art\xspace}
\newcommand{\sota}{SoTA\xspace}
\newcommand{\dpplong}{dynamic programming\xspace}
\newcommand{\dpp}{DP\xspace}

\newcommand{\ont}{ONT\xspace}

\newcommand{\dna}{DNA\xspace}
\newcommand{\gpulong}{Graphics Processing Unit\xspace}
\newcommand{\gpu}{GPU\xspace}
\newcommand{\fpgalong}{Field-Programmable Gate Array\xspace}
\newcommand{\fpga}{FPGA\xspace}
\newcommand{\cpulong}{Central Processing Unit\xspace}
\newcommand{\cpu}{CPU\xspace}

\newcommand{\asic}{ASIC\xspace}
\newcommand{\hslong}{hardware/software co-designed\xspace}
\newcommand{\hs}{HW/SW co-designed\xspace}
\newcommand{\mllong}{Machine Learning\xspace}

\newcommand{\bioinformatics }{Bioinformatics\xspace}

\newcommand{\salong}{sense amplifier\xspace}
\newcommand{\sa}{SA\xspace}
\newcommand{\tcamlong}{ternary content addressable memory\xspace}

\newcommand{\tcam}{TCAM\xspace}
\newcommand{\fifolong}{first in first out\xspace}
\newcommand{\fifo}{FIFO\xspace}
\newcommand{\fsmlong}{finite-state machine\xspace}
\newcommand{\fsm}{FSM\xspace}

\newcommand{\cmos}{CMOS\xspace}

\newcommand{\reramlong}{resistive random-access memory\xspace}
\newcommand{\reram}{ReRAM\xspace}

\newcommand{\scoutinglogic}{Scouting Logic\xspace}

\newcommand{\ambit}{Ambit\xspace}
\newcommand{\simdram}{SIMDRAM\xspace}

\newcommand{\bplong}{base-pair\xspace}
\newcommand{\bp}{bp\xspace}
\newcommand{\kbp}{Kbp\xspace}
\newcommand{\pacbio}{PacBio\xspace}

\newcommand{\clr}{CLR\xspace}

\newcommand{\teslagpu}{Tesla-K80\xspace}
\newcommand{\intelxeonqce}{Intel® Xeon® CPU E5-2680\xspace}

\newcommand{\edlib}{Edlib\xspace}

\newcommand{\pbsimthree}{PBSIM3\xspace}

\newcommand{\parasail}{Parasail\xspace}

\newcommand{\onttenk}{ONT\_10K\xspace}
\newcommand{\onthunk}{ONT\_100K\xspace}

\newcommand{\roilong}{region of interest\xspace}
\newcommand{\roi}{ROI\xspace}

\newcommand{\gatexorop}{XOR\xspace}
\newcommand{\gateandop}{AND\xspace}
\newcommand{\gateorop}{OR\xspace}

\newcommand{\minimap}{Minimap2\xspace\xspace}
\newcommand{\sneakysnake}{SneakySnake\xspace}
\newcommand{\fpgasneakysnake}{Snake-on-Chip\xspace}
\newcommand{\gpusneakysnake}{Snake-on-GPU\xspace}

\newcommand{\lzclong}{leading zero count\xspace}
\newcommand{\lzc}{LCZ\xspace}

\newcommand{\tnlong}{True Negative\xspace}
\newcommand{\tn}{TN\xspace}
\newcommand{\tplong}{True Positive\xspace}
\newcommand{\tp}{TP\xspace}
\newcommand{\fnlong}{False Negative\xspace}
\newcommand{\fn}{FN\xspace}
\newcommand{\fplong}{False Positive\xspace}
\newcommand{\fp}{FP\xspace}

\newcommand{\geneguardianlongreadalg}{LongGeneGuardian\xspace}

\newcommand{\filterfuselongreadacc}{FilterFuse\xspace}

\newcommand{\rattlesnakejakesamos}{RattlesnakeJake\xspace}

\newcommand{\shd}{SHD\xspace}

\newcommand{\counttcam}{Count-TCAM\xspace}
\newcommand{\patterndetect}{Pattern-Detect\xspace}
\newcommand{\outputselect}{Output-Select\xspace}
\newcommand{\pdtcam}{PD-TCAM\xspace}
\newcommand{\ostcam}{OS-TCAM\xspace}

\newcommand{\alignercpualignment}{\emph{$Edlib_{CPU}$-Alignment}\xspace}
\newcommand{\sscpufilter}{\emph{$SS_{CPU}$-Filter}\xspace}
\newcommand{\sscpualignment}{\emph{$SS_{CPU}$-Alignment}\xspace}

\newcommand{\filterfusereram}{\emph{$FilterFuse_{\reram}$}\xspace}
\newcommand{\filterfusecmos}{\emph{$FilterFuse_{\cmos}$}\xspace}
\newcommand{\rattlesnakejakereram}{\emph{$RattlesnakeJake_{\reram}$}\xspace}
\newcommand{\grimthreedstacked}{\emph{$GRIM_{3DStacked}$}\xspace}
 
\newcommand{\timeImprovementFilteringGeneguardianlongreadalgOverSneakysnakeCPUUpto}{120.47$\times$\xspace} 
\newcommand{\timeImprovementEndtoendFilterfuselongreadaccOverEdlibUpto}{5207.63$\times$\xspace} 
\newcommand{\timeImprovementEndtoendFilterfuselongreadaccOverSneakysnakeCPUUpto}{49.14$\times$\xspace} 
\newcommand{\timeImprovementEndtoendFilterfuseReRAMlongreadaccOverFilterfuseCMOSUpto}{28.93\%\xspace} 
\newcommand{\alignmentContributionEndtoendFilterfuseLongreadUpto}{90.10\%\xspace}
\newcommand{\alignmentContributionEndtoendFilterfuseLongreadMinimum}{59.67\%\xspace}
\newcommand{\powerReductionFilterfuseLongreadVsGpusneakysnake}{79.7\%\xspace}

\newcommand{\timeImprovementEndtoendSneakysnakeCPUOverEdlibUpto}{158.76$\times$\xspace} 
\newcommand{\minDatamovementOverheadGPUSneakysnakeLongReads}{84.56\%\xspace}
\newcommand{\maxDatamovementOverheadGPUSneakysnakeLongReads}{98.8\%\xspace}

\newcommand{\affilTUD}[0]{\textsuperscript{\S}}

\newcommand{\affilETH}[0]{\textsuperscript{$\dagger$}}

%% file: Sections/00_abstract.tex
\begin{abstract}

With the industry moving towards sequencing of accurate long reads (as they favor accurate and more efficient reconstruction of \dna), finding solutions that support efficient analysis of these reads becomes more necessary. The long execution time required for sequence alignment of long reads negatively affects genomic studies relying on sequence alignment. Although pre-alignment filtering as an extra step before alignment was recently introduced to mitigate sequence alignment for short reads, these filters do not work as efficiently for long reads. Moreover, even with efficient pre-alignment filters, the overall end-to-end (i.e., filtering + original alignment) execution time of alignment for long reads remains high, while the filtering step is now a major portion of the end-to-end execution time.

Our paper makes three contributions. First, it identifies data movement of sequences between memory units and computing units as the main source of inefficiency for pre-alignment filters of long reads. This is because although filters reject many of these long sequencing pairs before they get to the alignment stage, they still require a huge cost regarding time and energy consumption for the large data transferred between memory and processor. Second, this paper introduces an adaptation of a short-read pre-alignment filtering algorithm suitable for long reads. We call this \geneguardianlongreadalg. Finally, it presents \filterfuselongreadacc as an architecture that supports \geneguardianlongreadalg inside the memory. \filterfuselongreadacc exploits the \cimlong computing paradigm, eliminating the cost of data movement in \geneguardianlongreadalg.

Our evaluations show that \filterfuselongreadacc improves the execution time of filtering by \timeImprovementFilteringGeneguardianlongreadalgOverSneakysnakeCPUUpto for long reads compared to \sotalong (\sota) filter, \sneakysnake. \filterfuselongreadacc also improves the end-to-end execution time of sequence alignment by up to \timeImprovementEndtoendFilterfuselongreadaccOverSneakysnakeCPUUpto and \timeImprovementEndtoendFilterfuselongreadaccOverEdlibUpto compared to \sneakysnake with \sota aligner and only \sota aligner, respectively.

\end{abstract}

%% file: Sections/01_introduction.tex
\section{Introduction} 
\label{sec:introduction}

Sequence alignment is a pivotal process in genomics studies identifying similarities and differences in DNA, RNA, or protein sequences. By highlighting conserved regions and mutations, sequence alignment provides profound insights into the molecular function and evolution~\cite{needleman1970general-dynamicProgammingNeedlemanWunsch-153FromTechnologyDictatesAlgorithms, smith1981identification-dynamicProgammingSmithWaterman-152FromTechnologyDictatesAlgorithms}.  \sota sequence aligners employ \dpplong-based (\dpp) algorithms to achieve high accuracy with the cost of long latencies, low throughput, and energy inefficiencies, particularly when applied to long DNA sequences. These limitations directly affect the medical studies that benefit from sequence alignment.

Long reads and short reads are two types of sequencing reads used as inputs in the sequence alignment and are produced by different sequencing technologies~\cite{quail2008large-Illumina, segerman2020most-Illumina, fox2014accuracy-Illumina, rhoads2015pacbio-PacBio, ono2013pbsim-Old-PacBio-CLR-CCS-100Kbp-30Kbp, hu2021next-OverviewNextGenerationSequencing-new-PacBio-CLR-CCS-100Kbp-30Kbp, jain2016oxford-gagan28}. These two types of reads differ in their length, error rate (random errors due to technology used in obtaining them) or accuracy, application, usability, and cost. Overall, both long and short reads have their strengths and weaknesses, and the choice of sequencing technology depends on the specific research question and experimental design. However, although long-read sequence alignment faces several challenges (e.g., high error rates in long-read technologies, computational complexity due to longer read lengths, and issues with reference genome bias and uniqueness), currently, the industry is moving towards long reads~\cite{hu2021next-OverviewNextGenerationSequencing-new-PacBio-CLR-CCS-100Kbp-30Kbp, NanoBLASTer-Nanopore, Haghshenas2018lordFASTSA-Pacbio, BLASR}. This is because of the ability of long-read sequence alignment to resolve complex genomic regions, identify structural variations, and aid in epigenetic studies. Therefore, devising algorithms and/or hardware that can accelerate long-read sequence alignment while accurately mapping long reads to reference genomes and handling the unique characteristics of long-read datasets is of utmost importance in the coming years.

Previous works typically took two directions to address the inefficiency in sequence alignment for long reads~\cite{NanoBLASTer-Nanopore, Haghshenas2018lordFASTSA-Pacbio, YAHA-long_read_alignment, BLASR, li2016minimap-minimapandMiniasm-gagan45, rHAT_alignment_long_read, Sovic2016FastAS-Fast_sensitive_mapping_GraphMap}. First, some works simplified the process using better sketching or chaining algorithms or heuristics for \dpplong (\dpp) part of the general alignment algorithms~\cite{li2016minimap-minimapandMiniasm-gagan45, rHAT_alignment_long_read, Sovic2016FastAS-Fast_sensitive_mapping_GraphMap}. This backtracking step involves irregular memory access patterns that are challenging for hardware implementation. Second, a few works~\cite{SneakySnake, kim2018grim-grimfilterprealignment} propose a filtering step before alignment, called pre-alignment filtering\footnote{We use the term filter and pre-alignment filter interchangeably hereafter.}, to significantly speed up the end-to-end sequence alignment of (long) reads by heuristically replacing the need for expensive \dpp solutions for many inputs in the first place. These filters use a pre-defined edit distance threshold between the inputs and quickly determine whether or not an alignment (i.e., \dpp) should be granted. \sota pre-alignment filters~\cite{SneakySnake} speed up the sequence alignment so much so that they themselves become the (next) bottleneck in the end-to-end sequence alignment procedure. Therefore, there is a need for a more efficient design to tackle the filtering bottleneck in the sequence alignment pipeline of long reads.

We identify four shortcomings in pre-alignment filtering for target long reads. First, there is currently only one single filter, \sneakysnake~\cite{SneakySnake}, that supports pre-alignment filtering for long reads. Second, only the \cpulong{}s (\cpu{}s) implementation of \sneakysnake supports long reads, although \sneakysnake accelerates the pre-alignment filters for short reads on \cpu{}s, \gpulong{}s (\gpu{}s), and \fpgalong{}s (\fpga{}s). This happens due to strict assumptions on data and heuristics on the \gpu and \fpga versions. Third, on both types of reads, even a \sota filter, e.g., \sneakysnake, becomes the new computational bottleneck when considering the end-to-end alignment process (i.e., filtering step + sequence alignment step)~\cite{SneakySnake, kim2018grim}. Fourth, data movement bottlenecks the performance of \sota pre-alignment filters, particularly for long reads. This means that filters spend more time on moving sequencings from memory units to processing units compared to the time they spend on performing the computations necessary for the filtering. This shortcoming is important because most of the sequence pairs that go to the pre-alignment filters turn out to be unnecessary and will be filtered out eventually~\cite{SneakySnake}. Therefore, there is a need for a design that can overcome these shortcomings and resolve the bottleneck by avoiding wasted work (i.e., time and energy consumption) caused by data movement in the system.

To this end, We propose \geneguardianlongreadalg, a lightweight and memory-friendly pre-alignment filtering algorithm that supports long reads and performs on par with \sota pre-alignment filters regarding accuracy metrics (\textbf{Contribution \#1}). \geneguardianlongreadalg makes no assumption on data arrangement in a memory crossbar, making it suitable for a \cim design. We then present \filterfuselongreadacc, a \hslong (\hs) accelerator based on \cimlong (\cim) that supports \geneguardianlongreadalg for long reads (\textbf{Contribution \#2}). The simplicity of \geneguardianlongreadalg and its minimum assumptions on data placement inside a conventional memory array makes it compatible with a restricted yet realistic \cim design. \filterfuselongreadacc architecture is memory/technology independent, i.e., the memory arrays can be of any memory technology, such as \dramlong (\dram) or \reramlong (\reram), as long as they support key operations such as logical \gatexorop and associative search. Finally, we comprehensively evaluate \geneguardianlongreadalg and \filterfuselongreadacc (\textbf{Contribution \#3}). Our results show that \geneguardianlongreadalg achieves an accuracy on par with \sneakysnake, the \sota filter, for the long-read filtering. Since \geneguardianlongreadalg does not introduce any extra false negatives in the filtering process and does not replace the sequence alignment, one can still employ \geneguardianlongreadalg with any sequence aligner. When accelerated with memristor-based memory components, \filterfuselongreadacc accelerates the filtering step by up to \timeImprovementFilteringGeneguardianlongreadalgOverSneakysnakeCPUUpto over \sneakysnake, for the same read long-read dataset. Our evaluations also show that the using \filterfuselongreadacc for pre-alignment filtering accelerates the end-to-end alignment by up to \timeImprovementEndtoendFilterfuselongreadaccOverSneakysnakeCPUUpto and \timeImprovementEndtoendFilterfuselongreadaccOverEdlibUpto, compared to the case of \sneakysnake{}+long-read aligner and a standalone long-read aligner, respectively.

%% file: Sections/02_background.tex
\section{Background} 
\label{sec:background_and_relatedWork}

In this section, we briefly discuss the necessary background for our work. For more details, we refer the readers to review papers on the same topics~\cite{alser2022molecules-FrommoleculestogenomicvariationsAcceleratinggenomeanalysis, pages2022comprehensive-gagan30, hamdioui2015memristor}.

\subsection{Sequence Alignment} 
\label{subsec:sequence_alignment-background_and_relatedWork}

Read mapping is a common step in many genome analysis studies. Read mapping is needed to determine where a read originates from. For this purpose, one must compare the read to every possible location of a reference genome for the organism's species. Due to the sheer size of the reference genome (\textgreater{}3 billion \bplong{}s (\bp{}s), i.e., characters of \{A, C, G, T\}, for the human genome), this would be a computationally prohibitively intensive task. To add to that, the reads might contain edits that further complicate this process. Edits are defined as the differences between two strings and can be substitutions, insertions, and deletions~\cite{Zook2020ARB,firtina2021blend,li2018minimap2}. Currently, most potential solutions for this computationally intensive task follow the seed-and-extend approach~\cite{FastHASH-Xin2013}.

The seed-and-extend has three steps: (1) indexing: an offline, preparatory step undertaken on a pre-identified reference genome, (2) seeding: a step to employ the index structure to identify possible mapping points for each read within the reference genome, using smaller segments; i.e., seeds, from every read, and (3) sequence alignment: a step using dynamic programming algorithm to map candidates to decide the most appropriate mapping points in the reference for the input read.

\sota sequence aligners use computationally costly \dpp algorithms in their traceback step to prevent unnecessary, duplicate work. Here, a large problem is split into smaller sub-problems solved recursively by applying the notion of divide-and-conquer. Previous works~\cite{lassmann2005kalign, crochemore2003subquadratic, fei2018fpgasw, luo2013soap3, bu2013hsa-AlignmentApproximateAcceleration} suggested two main directions to improve the sequencing alignment step itself: (1) accelerating the \dpp algorithms directly or using heuristics, and (2) exploiting the inherent parallelism of algorithms and accelerating them using high-performance computing platforms, e.g., \fpga{}s and \gpu{}s.

\subsection{Pre-Alignment Filtering} \label{subsec:prealignment_filtering-background_and_relatedWork}

\begin{figure}[htbp]
\centering
    \includegraphics[width=1\linewidth]   {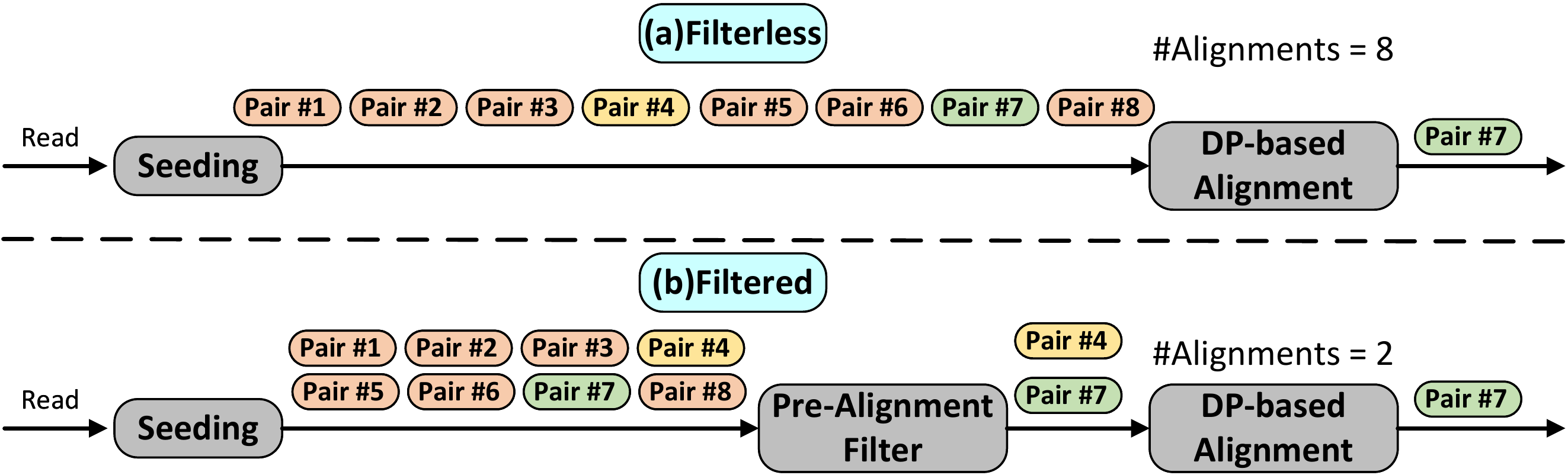}
    \caption{The concept of pre-alignment filtering.}
    \label{fig:pre_alignment_filtering_concept-prealignment_filtering-background_and_relatedWork}
\end{figure}

A recent, different approach to mitigate the alignment cost is pre-alignment filtering~\cite{alser2017gatekeeper, alser2019shouji, SneakySnake, kim2018grim}, which aims to reduce the number of pairings that must be evaluated by alignment. Filters achieve this by approximating the edit distance between the read and reference and removing pairings with an edit distance that greatly exceeds the alignment threshold, as demonstrated in \fig{\ref{fig:pre_alignment_filtering_concept-prealignment_filtering-background_and_relatedWork}}. This is effective as most pairings generated by the seeding process are dissimilar (contain many edits), while there exist only a few accepted mappings~\cite{xin2015shifted-SHD2015, alser2017gatekeeper, alser2019shouji, SneakySnake, kim2018grim}.

Pre-alignment filters perform the edit-distance estimation using less computationally demanding algorithms, thereby decreasing the end-to-end execution time of alignment (i.e., filtering time + alignment time). An example of these less computationally demanding algorithms is Hamming distance, measuring the number of differing positions between two equal-length strings, often DNA or RNA sequences. Using pre-alignment filters is orthogonal to the acceleration of alignment algorithms. Therefore, improvements to one can be combined with those of the other to achieve even greater benefits.

Filtering accuracy is typically evaluated based on 4 rates~\cite{SneakySnake, kim2018grim, alser2019shouji}: \tplong (\tp), \tnlong (\tn), \fplong (\fp), and \fnlong (\fn) rates. \tp is the ratio of pairs that filter correctly accepts/passes as they require \dpp-based alignment. The higher the \tp, the better. \fp is the ratio of pairs that filter incorrectly accepts/passes even though they do not require \dpp-based alignment. The lower the \fp, the better, as we would spend less time and cost for the alignment.

\subsection{\cimlong (\cim)} 

\label{subsec:subsec:pimcim_memristors-background_and_relatedWork}

The recently re-ignited \cimlong (\cim)\footnote{Interchangeably also known as \pimlong (\pim).} paradigm~\cite{Cellular_Logic_in_Memory1969-PIMCIM5decades, Logic_in_Memory1970-PIMCIM5decades, shaw1981non-NonVonDatabase-PIMCIM5decades, Computational_Ram1970-PIMCIM5decades, EXECUBE-A1994-PIMCIM5decades, Terasys1995-PIMCIM5decades, intelligent_RAM1997-PIMCIM5decades, Active_Pages_intelligent_Memory1998-PIMCIM5decades, FlexRAM1999-PIMCIM5decades, ProgrammingFlexRAM2003-PIMCIM5decades, DIVA2002-PIMCIM5decades, Smart_memories2000-PIMCIM5decades} is a promising way to minimize data movement and saves us time and energy as the bandwidth is the largest near the memory. Previous works~\cite{chi2016prime, ankit2019puma, shahroodi2022krakenonmem} from industry and academia propose architectures based on this paradigm to improve the performance and energy consumption in applications with relatively small and/or simple computations and work on large amounts of data, such as those in \mllong and \bioinformatics. We argue that pre-alignment filtering algorithms enjoy the same properties.

A \cim design can use different memory technologies, e.g., established ones like DRAM~\cite{seshadri2017ambit, li2017drisa, gao2019computedram, ferreira2022pluto} and SRAM~\cite{aga2017computecache, fujiki2019duality, li2016pinatubo, eckert2018neural}, or emerging ones such as memristors~\cite{xie2017scouting, zahedi2020efficient}. We evaluate our proposal using \dram, \sram, and memristors as our underlying technology in \cim due to the maturity of literature on them and our access to real \reram-based chip measurements and accurate memory models for them. We leave an even more comprehensive evaluation of our design based on different technologies for future work.

%% file: Sections/03_motivation.tex
\section{Motivation} 
\label{sec:motivation}

\subsection{Importance of Sequence Alignment}
\label{subsec:importance_sequence_alignment_vs_sequencing-motivation}

For two key reasons, the acceleration of sequence alignment (and pre-alignment filtering) for long reads remains an important research direction. First, data reuse: sequencing is a one-off task, but its output (genomic data) is analyzed repeatedly. With over 29 peta bases of data available as FASTQ files on the SRA database~\cite{SRA-NCBI-29PetaBases}, in many of the non-real-time use cases, the focus shifts from the cost of sequencing itself to the speed of the alignment. This vast amount of data is often reused, making alignment throughput crucial. Second, analysis bottleneck: For comprehensive genomic studies using ultra-long (\ont) and accurate long reads (\pacbio)~\cite{alser2022molecules-FrommoleculestogenomicvariationsAcceleratinggenomeanalysis}, i.e., our target reads in this work, the speed of modern sequencers outpaces subsequent analyses like sequence alignment, as benchmarked in previous works~\cite{alser2022molecules-FrommoleculestogenomicvariationsAcceleratinggenomeanalysis}. This makes alignment, not sequencing, the bottleneck in many (non-real-time) cases and, therefore, worthy of acceleration, even in the hardware.

\subsection{Long Reads vs. Short Reads}
\label{subsec:long_reads_vs_short_reads-motivation}

Different sequencing technologies produce reads with various features regarding accuracy and read length. The accuracy of a sequencing machine is described as the percentage of base pairs it has correctly extracted from a \dna sample. The read length is the number of base pairs that constitute it. These two metrics determine the alignment threshold (\sect{\ref{subsec:prealignment_filtering-background_and_relatedWork}}) and the edit margin, also known as the \roilong (\roi), in the evaluation of pre-alignment filters\footnote{For long reads, the \roi is 2-7\% of the read length~\cite{alser2017magnet, alser2019shouji, hach2010mrsfast}.}.

The read accuracy matters as sequencing aims to compare samples of \dna to find differences/similarities between them. For this, we need to be able to differentiate between actual differences between the samples and sequencing errors. Sequencing errors are defined as differences between the extracted read and the \dna sample. These errors originate from deficiencies in sequencing technologies. Having highly accurate reads is favorable.

For two main reasons, it is favored to have accurate long reads as long as the length does not (significantly) hurt the accuracy. First, accurate long reads simplify the reconstruction of the original \dna compared to separate shorter reads. Like fitting pieces of a puzzle together, it is easier to do this with fewer long reads than many short reads. Second, the probability of a short read aligning with multiple parts of a reference genome is much higher than is the case with long reads. Consequently, finding the source of a mutation in the \dna is much harder with short reads.

Therefore, although short reads remain popular due to their availability in genomics libraries, the industry will require developing techniques to process accurate, long reads.

\subsection{Limitations of \sota filters for long reads}
\label{subsec:Limitations_sota_pre_alignment_filters_long_reads-motivation}

\textbf{Support for long reads.}
Although previous works~\cite{alser2017magnet, alser2019shouji, xin2015shifted-SHD, kim2018grim, SneakySnake} propose various methods of pre-alignment filters, their methods rarely work on both types of reads due to underlying assumptions on input data or how to determine/approximate the similarity of two sub-strings. Specifically, among all the proposed filters, only \sneakysnake supports long reads.

\noindent
\textbf{Filter is the new bottleneck.}
We profile \sneakysnake as the filter and \parasail as a sequence aligner over different percentages of edit distances and datasets. We refer to \sect{\ref{sec:experimental_setup_and_evaluation_methodology}} for details on our evaluation methodology.

\begin{figure*}[hbtp] 
\centering
    \includegraphics[width=0.8\linewidth]{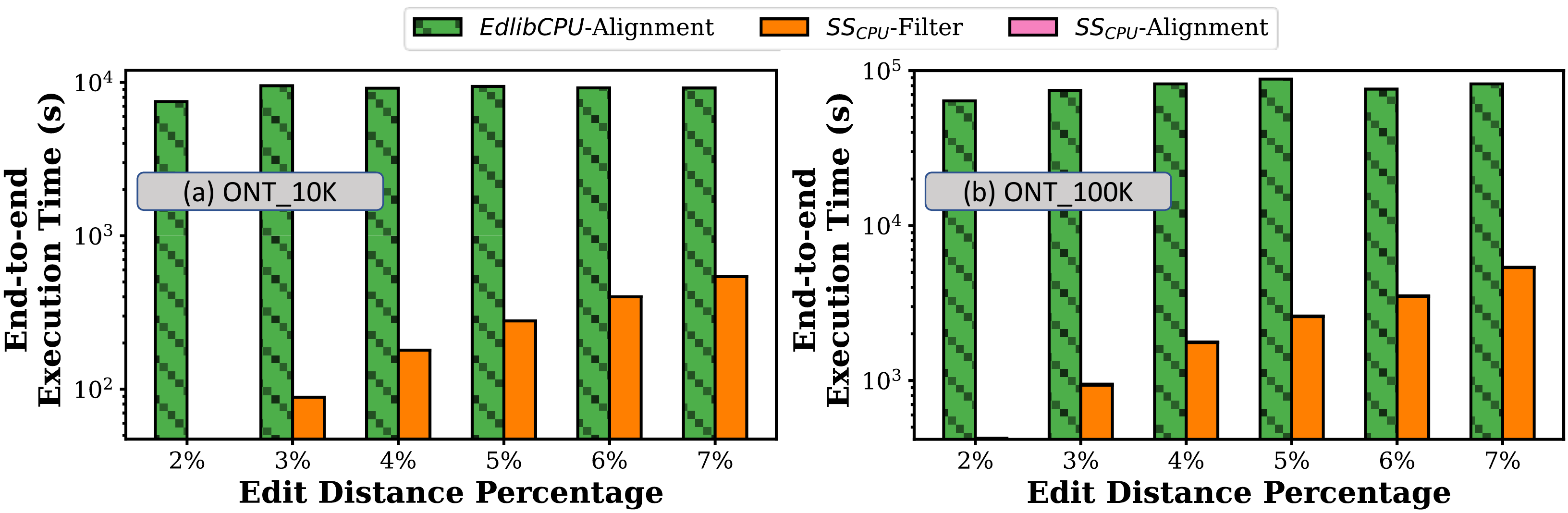}
    \caption{Contribution of \sneakysnake to the end-to-end execution time for long reads.}
    \label{fig:partial_EndtoEnd_ExecutionTime_longRead_PacBio9810KAndPacBio98100K_Alignment_SneakySnakeCPUAndAlignment-long_reads_vs_shor_reads-motivation}
\end{figure*}

\fig{\ref{fig:partial_EndtoEnd_ExecutionTime_longRead_PacBio9810KAndPacBio98100K_Alignment_SneakySnakeCPUAndAlignment-long_reads_vs_shor_reads-motivation}} presents the execution time for end-to-end alignment for our two representative datasets. The \alignercpualignment shows the time it takes to perform alignment on \cpu using a \sota aligner. The \sscpufilter and \sscpualignment present the time required for filtering and then alignment of unfiltered read-reference pairs, respectively, using the \sota filter and aligner on \cpu. For better readability, we limit the results to the edit thresholds in the appropriate \roi (see \sect{\ref{subsec:long_reads_vs_short_reads-motivation}}), similar to \sneakysnake~\cite{SneakySnake}. The y-axis is in logarithmic scale.

\begin{figure*}[htbp] 
\centering
    \includegraphics[width=0.7\linewidth]{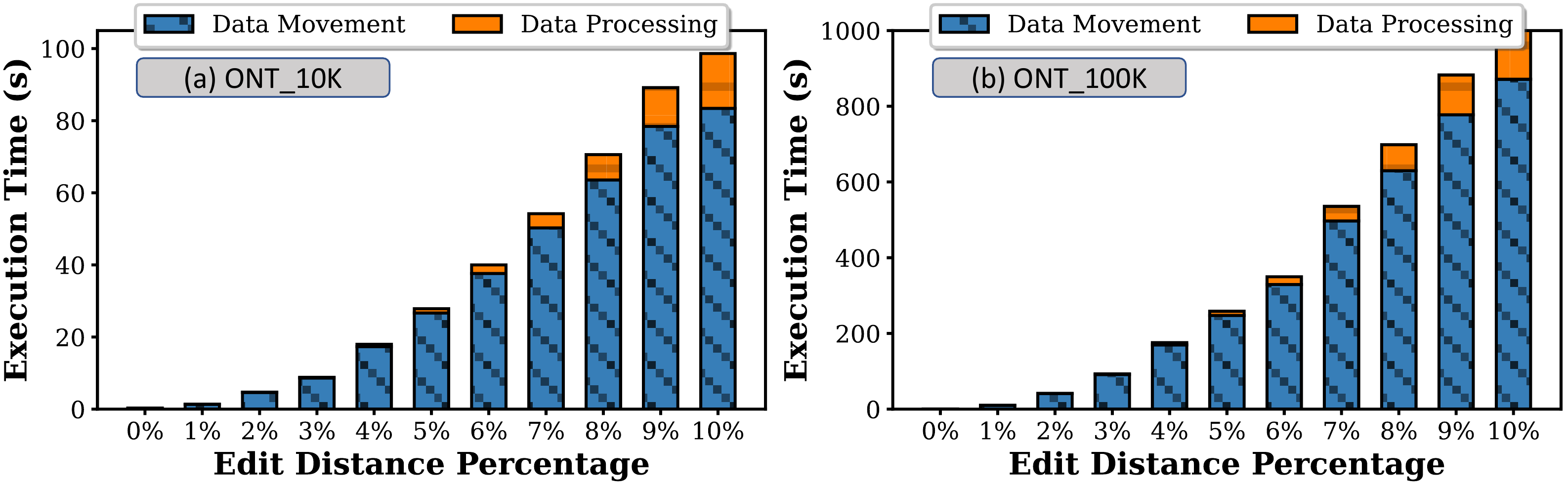}
    \caption{Contribution of data movement in long-read filtering.}
    \label{fig:datamovement_bottleneck_longRead_PacBio9810KAndPacBio98100K_SneakySnakeGPU-long_reads_vs_shor_reads-motivation}
\end{figure*}

We make two main observations. First, a pre-alignment filter improves the end-to-end execution time by up to \timeImprovementEndtoendSneakysnakeCPUOverEdlibUpto. This shows that this filtering is effective for long reads. Second, the majority of the new end-to-end execution time is spent on filtering rather than alignment. This majority is high enough to the extent that the alignment time is almost invisible compared to the filtering time, even with a logarithmic scale. This shows that to improve the end-to-end execution time, the filtering step is the new bottleneck to focus on.

\noindent
\textbf{Data movement limits accelerated filters.}
\fig{\ref{fig:datamovement_bottleneck_longRead_PacBio9810KAndPacBio98100K_SneakySnakeGPU-long_reads_vs_shor_reads-motivation}} presents the breakdown of execution time for a \sota accelerated filter, \sneakysnake on \gpu (\gpusneakysnake). Since \gpusneakysnake does not support long read filtering, we broke down the sequences into smaller non-overlapping chunks (with some post-processings on the host side) to achieve a rough estimation of the contribution of data movement to the total execution time in a hardware accelerator. We believe that our estimation for filtering long reads using \gpusneakysnake for short reads provides a reliable snapshot of the data movement challenge and is indeed conservative. Effective long-read filters often discard more reads due to increased errors, amplifying the data movement cost. Therefore, this estimation offers a conservative yet insightful glimpse into the issue, even when considering powerful accelerators like \gpusneakysnake.

We observe that \gpusneakysnake spends a minimum of \minDatamovementOverheadGPUSneakysnakeLongReads and up to \maxDatamovementOverheadGPUSneakysnakeLongReads of its execution time just transferring data from memory to \gpu. This shows that data movement constitutes the major part of the execution time for an accelerated filter. Although \gpusneakysnake does not support long read filtering directly, this experiment highlights accelerators alone, even if they are the leading hardware accelerator based on \gpu as it is the case in \gpusneakysnake, cannot bypass the core inefficiency of today's filters: data movement. 
%

\noindent
\textbf{No hardware acceleration support for long reads.}
\sneakysnake~\cite{SneakySnake} currently holds the highest accuracy and lowest execution time among all the existing pre-alignment filters. It supports both \gpu{}-based and \fpga{}-based implementations, called \gpusneakysnake and \fpgasneakysnake, respectively. However, the open-sourced implementations of \gpusneakysnake and \fpgasneakysnake only support short reads due to some heuristics and hard-coded assumptions in their implementations. Therefore, to this day, there is no hardware accelerator for long-read pre-alignment filtering.

\noindent
\textbf{Why not \sneakysnake (\sota filter) on \cim?}
Unfortunately, the \sneakysnake is not \cim friendly for two reasons: (1) it requires support for \lzclong (\lzc) or at least flexible \#shifts in data, which is costly in memory arrays, and (2) it requires perfect positioning of data inside memory units in respect to the boundaries of memory tiles. Unfortunately, even the most hardware-friendly implementation of \sneakysnake, \fpgasneakysnake, introduces two main challenges for a \cim implementation:

\begin{itemize}[leftmargin = *]
    \item \fpgasneakysnake requires the computation of an entire chip-maze of each sub-problem. The horizontal dimension of this chip maze is dependent on the number of bases of each sub-problem, and the vertical dimension is dependent on the examined edit distance. While the maze size is manageable for short reads, the resources required to store the chip maze for long reads would be too large to implement in a memory tile (the granularity we can expect a \cim design work with). For example, for long reads that reach up to 100\kbp{}s and have edit-distance thresholds of up to 10\%, required to process \ont reads, the chip maze would have to store up to 10 thousand rows of data per tile. 

    \item \fpgasneakysnake takes several iterations to find the optimal path through the chip maze. This process is time-consuming and requires a relatively large number of lookup tables to be implemented inside memory. 
\end{itemize}

\takeNN{We need efficient accelerators for pre-alignment filters of long reads with an emphasis on eliminating data movement.}

%% file: Sections/05_algorithm.tex
\section{\geneguardianlongreadalg Algorithm} 
\label{sec:proposed_algorithm_SSCIM}

We propose \geneguardianlongreadalg, a lightweight algorithm designed for long-read pre-alignment filtering having (1) rigid data accessibility in a \cim-enabled design and (2) lack of control on the start of a reference sequence stored in memory in mind. \geneguardianlongreadalg draws inspiration from  \rattlesnakejakesamos~\cite{shahroodi2023RattlesnakeJake}\footnote{\rattlesnakejakesamos~\cite{shahroodi2023RattlesnakeJake} is a recent and less accurate algorithm,  designed for short-read pre-alignment filtering.} and \fpgasneakysnake. However, \geneguardianlongreadalg differs from both \rattlesnakejakesamos and \fpgasneakysnake on how it interprets and processes its sub-problems, as discussed next.

\alg{\ref{alg:LongGeneGuardian_Algorithm-FilterFuse_algorithm-proposal_and_architecture}} presents \geneguardianlongreadalg, which as inputs takes the read-reference sequence pair, the number of permissible edits between input read and the reference sequence ($E$), two user parameters for (1) \#duplications in the input to mimic the shift behavior ($S$), and (2) the segment size ($T$). The segment is the simpler sub-problem granularity that \geneguardianlongreadalg works on. \geneguardianlongreadalg (\alg{\ref{alg:LongGeneGuardian_Algorithm-FilterFuse_algorithm-proposal_and_architecture}}) is based on three key observations:

\begin{algorithm}
\begin{algorithmic}[1]
\small \scriptsize
\algrenewcommand\algorithmicrequire{\textbf{Input:}}
\algrenewcommand\algorithmicensure{\textbf{Output:}}
\Require{Read, Reference, E, ReadLength, T, S}
\Ensure{Accept}
\State $N_{segment} \gets \lceil ReadLength/k \rceil$
\State $Matches \gets 0$
\State $S_{left} \gets 0$
\For{$i \in \{0: N_{segment}-1\}$}
    \State $Match \gets 0$
    \For{$e \in \{-E: +E\}$}
        \If{$|e| \ge S$}
            \State $S_{left} \gets S$
        \EndIf
        \If{$e > 0$}
            \State $S_{left} \gets - S_{left}$
        \EndIf
        \State $S_{right} = |e| - S_{left} $ 
        \State $ReadSegment \gets Read[i\times T+S_{left}: (i+1)\times T-1+S_{left}] $ 
        \State $ReferenceSegment \gets Ref[i\times T+S_{right}: (i+1)\times T-1+S_{right}]$
        \If{$ReadSegment == ReferenceSegment$}
            \State $Match \gets 1$
        \EndIf
    \EndFor
    \State $Matches \gets Matches + Match$
\EndFor
\State $Accept \gets (Matches >= N_{segment} - E)$
\State\Return $Accept$
\end{algorithmic}
\caption{\geneguardianlongreadalg Algorithm}
\label{alg:LongGeneGuardian_Algorithm-FilterFuse_algorithm-proposal_and_architecture}
\end{algorithm}

\begin{enumerate}[leftmargin = *]
    \item If two strings differ by $e$ edits, then all non-erroneous characters of the strings can be aligned in at most $e$ shifts.
    \item If two strings differ by $e$ edits, then they share at most $e+1$ identical sections.
    \item When you are interested in comparing two strings where you can shift one to the left by up to $e_{total}$ shifts while keeping the other one fixed, you can achieve the same results by shifting the first one to the left by $s_{left}$ and the second string to the right by $S_{right} = e_{total} - S_{left}$. 
    
\end{enumerate}

Exploiting these observations, \geneguardianlongreadalg divides the problem of performing Hamming distances on these shifted versions into the simpler exact-matching sub-problems of length $T$. \geneguardianlongreadalg first sets the \#segments to evaluate (Line 1 in \alg{\ref{alg:LongGeneGuardian_Algorithm-FilterFuse_algorithm-proposal_and_architecture}}) and initializes the necessary variables (Lines 2 and 3). After that, \geneguardianlongreadalg solves these sub-problems independently (Lines 5-19) and aggregates their results to find a lower bound for the number of exact matches between segments (Line 20). During this phase, \geneguardianlongreadalg creates $2e+1$ shifted segments to account for $e$ shifts to the left and right (Line 6). Then, \geneguardianlongreadalg performs the exact-matching check on the segments (Lines 14-16). Finally, \geneguardianlongreadalg determines whether or not it should accept the read-reference pair based on the total matches found and acceptable threshold (Lines 22).

\geneguardianlongreadalg detects edits in the segments using this intuition: if the section of the read processed in one sub-problem contains no edits, at least one of the segments must be free of errors. This means that \geneguardianlongreadalg can check whether any of the Hamming distances belonging to the segment contains only `0's. This allows the detection of \#segments without edits. By subtracting this from the total \#segments, \geneguardianlongreadalg finds \#segments that do contain errors.

\geneguardianlongreadalg and \rattlesnakejakesamos differ in that \geneguardianlongreadalg exploits observation \#3 unlike \rattlesnakejakesamos. Therefore, unlike \rattlesnakejakesamos, which keeps the read untouched and only shifts the reference, \geneguardianlongreadalg creates the shifted sub-sequences by shifting both read and reference segments to left and right. This change is necessary for long reads because otherwise, we incur unacceptable overhead for storing the shifted reference sequences of our long reads in the memory/hardware when it comes to our accelerator. We elaborate on this further in \sect{\ref{subsec:long_read_compatibility_discussion-proposed_architecture_hardware_accelerator_SSCIM}}.

\geneguardianlongreadalg and \fpgasneakysnake both divide the problem of performing Hamming distances on the shifted versions of sequences into smaller sub-problems. However, \geneguardianlongreadalg differs from \fpgasneakysnake in that it does not count the number of edits in each segment but detects any edit's presence. In doing so, the results of the two algorithms only differ if multiple errors occur in the same part of the read. This causes \geneguardianlongreadalg to accept some read-reference pairs (that could have been filtered) due to the abstraction of actual \#edits in segments. However, our evaluations using real datasets in \sect{\ref{sec:evaluations_and_results}} show that it is unlikely that two edits exist in the same segment when the segments are small. Moreover, each read contains a small \#edits relative to its read length. Therefore, the decrease in accuracy is still acceptable as \geneguardianlongreadalg can still distinguish true mappings (similar) from obviously false mappings (dissimilar) and provide enough speed-up (\sect{\ref{sec:evaluations_and_results}}).

A key advantage of \geneguardianlongreadalg's approach is that every exact matching operation corresponding to the different shifts can be efficiently processed independently. This removes the need to collect all Hamming distances to create the chip-maze and removes the iterative nature of the chip-maze traversal step. This and the segmentation into sub-problems make \geneguardianlongreadalg particularly suitable for \cim.

%% file: Sections/06_architecture_overview.tex
\section{\filterfuselongreadacc Architecture} 
\label{sec:proposed_architecture_hardware_accelerator_SSCIM}

We implement \geneguardianlongreadalg using \cim, called \filterfuselongreadacc. While \filterfuselongreadacc is designed to support long reads, it remains flexible and supports a wide range of data sets, edit-distance thresholds, and even short-reads filtering algorithms.

\subsection{\filterfuselongreadacc Overview}
\label{subsec:architecture_overview-proposed_architecture_hardware_accelerator_SSCIM}

\begin{figure}[htbp]
\centering
    \includegraphics[width=1\linewidth]{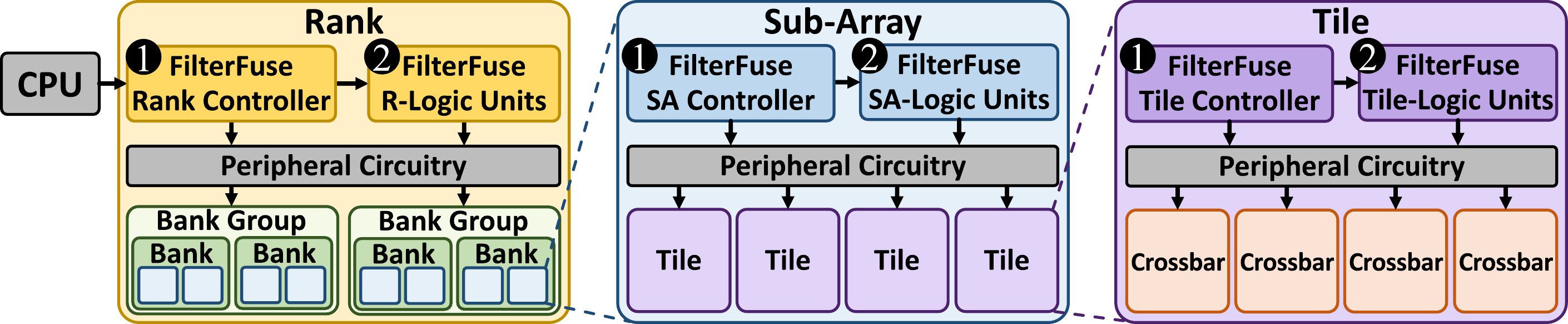}
    \caption{Overview of \filterfuselongreadacc.}
    \label{fig:filterFuse-architecture_overview-proposed_architecture_hardware_accelerator_SSCIM}
\end{figure}

\fig{\ref{fig:filterFuse-architecture_overview-proposed_architecture_hardware_accelerator_SSCIM}} presents an overview of the \filterfuselongreadacc.

\filterfuselongreadacc follows the typical memory hierarchies (i.e., bank groups, banks, sub-arrays, and tiles) found in conventional memories to improve resource utilization. However, \filterfuselongreadacc augments various hierarchy levels with small specialized controllers~\circled{1} and logic units~\circled{2} to enable the required operations in \geneguardianlongreadalg. Each level contains an \fsmlong-based (\fsm) controller that controls all the logic units and the lower-level controllers.

%% file: Sections/07_tile_architecture.tex
\subsection{Tile Architecture}
\label{subsec:tile_architecture-filterFuse-proposed_architecture_hardware_accelerator_SSCIM}

\begin{figure}[htbp]
\centering
    \includegraphics[width=0.9\linewidth]{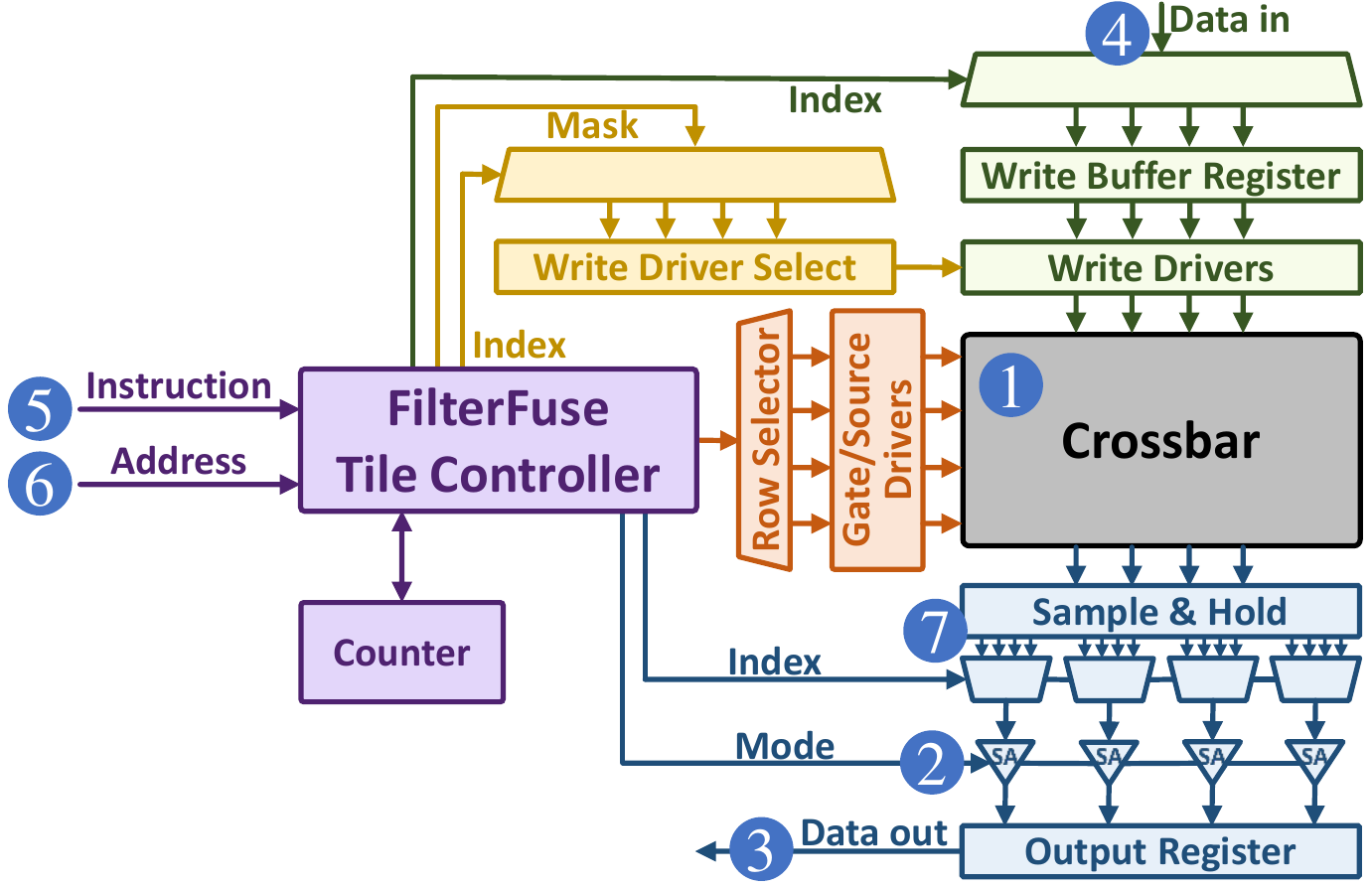}
    \caption{Overview of the tile architecture.}
    \label{fig:tile_architecture-filterFuse-proposed_architecture_hardware_accelerator_SSCIM}
\end{figure}

Tiles are the lowest and one of the most critical architectural levels in \filterfuselongreadacc. \fig{\ref{fig:tile_architecture-filterFuse-proposed_architecture_hardware_accelerator_SSCIM}} presents the architecture of a tile. 

Each tile is an array of memory cells forming a crossbar structure \Bcircled{1}. \filterfuselongreadacc supports any memory technology for its cell as long as it can support logical vector operation in the \sa. Peripheral circuits include multiple write drivers, \salong{}s (\sa{}s) \Bcircled{2}, row/column decoders, and multiplexers. The \sa{}s are modified for the required logical vector operation such as the required \gatexorop for the check required in Line 16 of \alg{\ref{alg:LongGeneGuardian_Algorithm-FilterFuse_algorithm-proposal_and_architecture}}. If someone uses \dram or memristor-based crossbars, the \sa{}s to support these operations are based on \ambit~\cite{seshadri2017ambit} and \scoutinglogic~\cite{li2016pinatubo, xie2017scouting}, respectively.

Each tile has an $n$-bit data output (\Bcircled{3}) and three inputs provided by the sub-array controller (besides the clock and reset signal): 
\begin{itemize}[leftmargin=*]
    \item The 'data-in' \Bcircled{4}: $n$ binary bits to be written to the crossbar. 
    
    \item The instruction signal \Bcircled{5}: to determine the behavior of the tile controller, selecting whether the tile should be idle, read, write, or perform an \gatexorop operation.

    \item The address signal \Bcircled{6}: to index the correct rows and columns of the crossbar to/from which the data should be written/read.
\end{itemize}

To execute \geneguardianlongreadalg, \filterfuselongreadacc first writes the read sequence to the appropriate memory locations. It then performs an \gatexorop between $n$-bits of the read and $n$-bits of the reference sequence (which is already written in the memory), where $n$ indicates the number of \sa{}s in the tile. Each of these $n$-bits \gatexorop within a tile contains the bit-wise operations required for the comparison in Line 16 of \alg{\ref{alg:LongGeneGuardian_Algorithm-FilterFuse_algorithm-proposal_and_architecture}}.

The peripheral components interface between the digital architecture and the (analog) crossbar. They also act as intermediate storage of input data to overcome the difference in timing between the read/write time of the crossbar and the clock period of other digital components.

If needed, in the case of memristor-based crossbars, for example, \filterfuselongreadacc implements interleaving within the column multiplexing logic. The tile architecture indexes a series of de-multiplexers at the output of the sample-and-hold circuit (\Bcircled{7} in \fig{\ref{fig:tile_architecture-filterFuse-proposed_architecture_hardware_accelerator_SSCIM}}) to select the correct memory rows. This way, the correct digital output of the \sa{}s is placed in an output register and can be accessed by the sub-array controller.

%% file: Sections/08_subarray_architecture.tex
\subsection{Sub-Array Architecture}
\label{subsec:subarray_architecture-filterFuse-proposed_architecture_hardware_accelerator_SSCIM}

\begin{figure}[htbp]
\centering
    \includegraphics[width=1\linewidth]{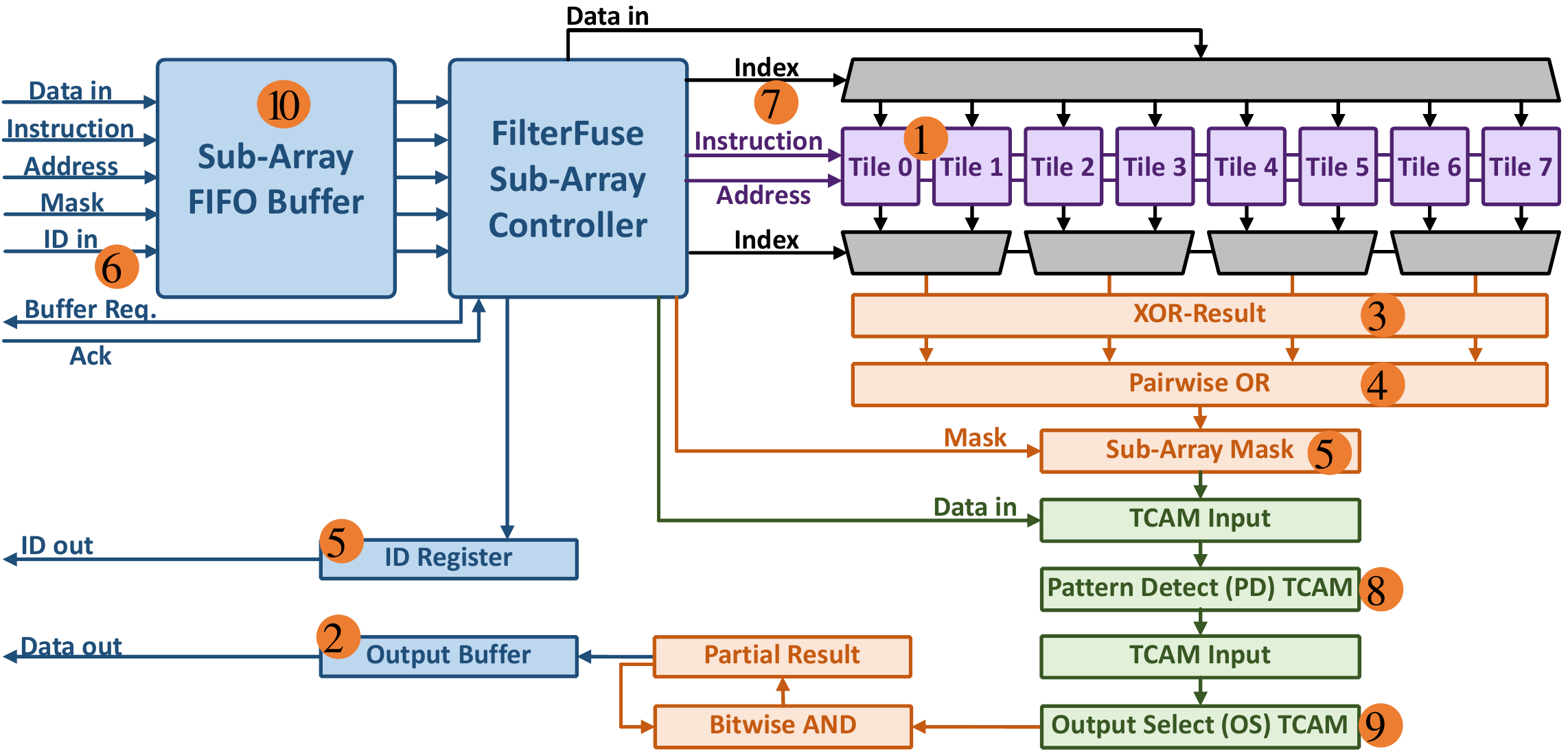}
    \caption{Overview of the sub-array architecture.}
    \label{fig:subarray_architecture-filterFuse-proposed_architecture_hardware_accelerator_SSCIM}
\end{figure}

\fig{\ref{fig:subarray_architecture-filterFuse-proposed_architecture_hardware_accelerator_SSCIM}} presents an overview of \filterfuselongreadacc's sub-array architecture.

The sub-arrays in \filterfuselongreadacc are the second main computational units, where multiple tiles (\Ocircled{1}) are grouped together to execute the complete logic in \geneguardianlongreadalg. Sub-arrays contain input and output buffers (\Ocircled{2} and \Ocircled{10}) required to reduce the stalling of the pipeline when the sub-array or output bus is occupied. Three main tasks of a sub-array are:

\begin{itemize}[leftmargin=*]
    
    \item Translating the results of bitwise \gatexorop from tiles (\Ocircled{3}) into base pair results (\Ocircled{4}). This is required to actually perform the comparison in Line 16 of \alg{\ref{alg:LongGeneGuardian_Algorithm-FilterFuse_algorithm-proposal_and_architecture}}.
    
    \item Performing masking on tile results that are not part of the read/reference pairings (\Ocircled{5}). This is required as not all of the tile-level results capture the target segments in \geneguardianlongreadalg.
    
    \item Tracking what read/reference pairing is being processed using an ID signal (\Ocircled{6}). This is required to actually perform the final checks in Line 22 of \alg{\ref{alg:LongGeneGuardian_Algorithm-FilterFuse_algorithm-proposal_and_architecture}} in the higher levels.
    
\end{itemize}

A sub-array takes seven steps to perform the necessary computation for a final bit-vector:
\begin{enumerate} [leftmargin=*]
    \item Passing the \gatexorop instruction to the tiles to perform the operation between the read sequence and one of the shifted reference sequences (\Ocircled{7}) (Line 22 of \alg{\ref{alg:LongGeneGuardian_Algorithm-FilterFuse_algorithm-proposal_and_architecture}}).
    
    \item Retrieving the results from the output buffers of the selected tiles by indexing a series of multiplexers. 
    
    \item Combining the results into an \gatexorop-result register (\Ocircled{3}).
    
    \item Performing a series of \gateorop to convert a bit-level \gatexorop result to a bit vector of base pair level result (\Ocircled{4}). 
    
    \item Masking parts of the \gateorop result that are not part of the actual read-reference pairing by \gateandop with a sub-array mask (\Ocircled{5}). 
    
    \item Querying a \tcamlong (\tcam), called \patterndetect \tcam (\pdtcam), with the masked results of the previous step (\Ocircled{8}).  
    
    \item Querying a second \tcam, called \outputselect \tcam (\ostcam), with the results of \pdtcam (\Ocircled{9}). 
\end{enumerate}

Note that the length of the operands at the sub-array level is determined by the minimum sub-problem size, i.e., $2*T$ (where $T$ is the segment size in \geneguardianlongreadalg) to account for the 2-bit encoding scheme commonly used in genomics accelerators~\cite{mansouri2022genstore, mao2022genpip, GenDP-ISCA2023-Das, fujiki2018genax}.

\filterfuselongreadacc uses \pdtcam and \ostcam to detect patterns of '1's and '0' in its input and construct one iteration of the final bit-vector based on patterns detected by the \pdtcam, respectively. This is necessary for \geneguardianlongreadalg and many previous pre-alignment filters. For example, \filterfuselongreadacc can use these \tcam{}s to detect the exact matches required in \geneguardianlongreadalg (Line 16 \alg{\ref{alg:LongGeneGuardian_Algorithm-FilterFuse_algorithm-proposal_and_architecture}}). \filterfuselongreadacc can use them to detect patterns of  “101” and “1001” required in \shd~\cite{xin2015shifted-SHD}. A similar method has been used by previous work~\cite{shahroodi2023RattlesnakeJake}.

\filterfuselongreadacc boosts filtering throughput by activating multiple sub-arrays in parallel to compute different read-reference pairings independently. To enable this parallel execution, \filterfuselongreadacc stores a different part of the reference on each sub-array and uses the input dataset to determine which sub-arrays are required for the computation.

It is possible that a read-reference pairing requires a sub-array that is still busy computing a different pairing, which creates contention over the sub-array, i.e., it stalls \filterfuselongreadacc until the sub-array is freed up. To combat this contention, each sub-array contains a \fifolong (\fifo) buffer (\Ocircled{10} in \fig{\ref{fig:subarray_architecture-filterFuse-proposed_architecture_hardware_accelerator_SSCIM}}) that stores all the input signals for the computation of a read-reference pairing until the sub-array finishes computing its current pairing. This eliminates the need to stall the rest of the pipeline until the \fifo buffer is full. Note that having a \fifo buffer, the order in which the computation of pairings finishes might differ from how they are supplied to the tiles. To prevent any issue, \filterfuselongreadacc assigns an ID to each pairing, which is also presented alongside the sub-array output (\Ocircled{2}). To prevent the potential contention over the output bus, which occurs when multiple sub-arrays finish their computation simultaneously, \filterfuselongreadacc uses a request-acknowledgment scheme, ensuring that outputs are read one after another.

%% file: Sections/09_banksandbankgrouns_architecture.tex
\subsection{Bank and Bank-Group Architecture}
\label{subsec:bank_and_bankgroup_architecture-filterFuse-proposed_architecture_hardware_accelerator_SSCIM}

\filterfuselongreadacc groups several sub-arrays as banks and then groups multiple banks into a bank group. This type of hierarchical structure is often found in conventional \dram-based technologies \cite{Kim2012ISCA_SALP_SubarraylevelparallelismDRAM}. \filterfuselongreadacc uses these levels to aggregate the results of intermediate steps in the target pre-alignment filtering algorithm. In the case of \geneguardianlongreadalg, \filterfuselongreadacc uses this step to perform the necessary aggregation of found matches (Line 20 of \alg{\ref{alg:LongGeneGuardian_Algorithm-FilterFuse_algorithm-proposal_and_architecture}}). Although these two levels in \filterfuselongreadacc have similar functionality, \filterfuselongreadacc adapts both for two main reasons. First, splitting the two levels reduces the fan-out of the required busses of each stage, which reduces the clock period required. Second, the multi-level approach improves the contention over the lower-level resources without adding excessive amounts of buffer overheads. These levels implement an acknowledgment scheme to determine the need for a stall due to the contention. Because the architecture consists of several layers, communication between the top level and the sub-array level happens over several clock cycles. The input buffers reduce this latency by providing an acknowledgment signal after only a single clock cycle.

%% file: Sections/10_rank_architecture.tex
\subsection{Rank Architecture}
\label{subsec:rank_architecture-filterFuse-proposed_architecture_hardware_accelerator_SSCIM}

\begin{figure}[htbp]
\centering
    \includegraphics[width=1\linewidth]{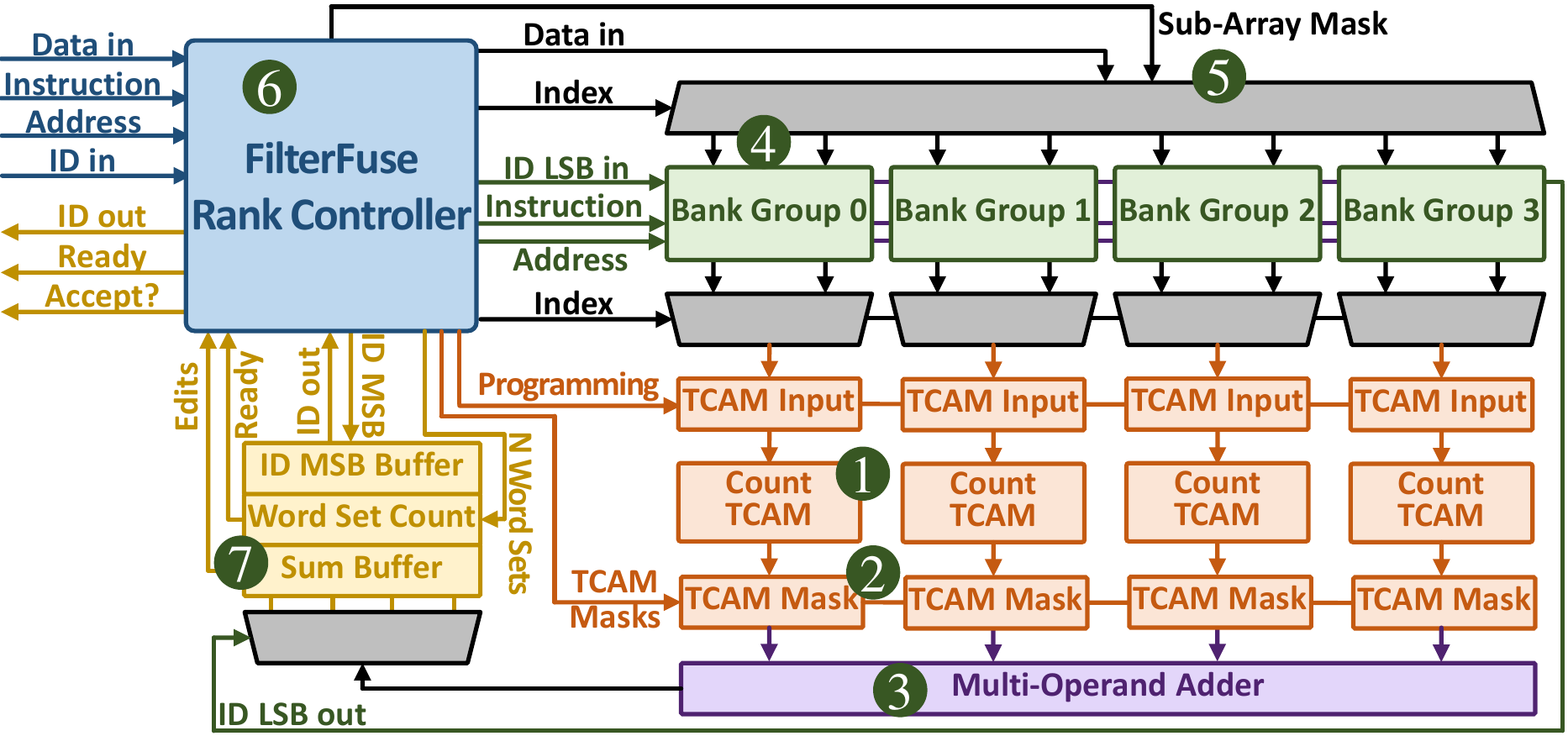}
    \caption{Overview of the rank architecture in \filterfuselongreadacc.}
    \label{fig:rank_architecture-filterFuse-proposed_architecture_hardware_accelerator_SSCIM}
\end{figure}

The rank level is the highest level of \filterfuselongreadacc that interfaces between the host device and \filterfuselongreadacc. \fig{\ref{fig:rank_architecture-filterFuse-proposed_architecture_hardware_accelerator_SSCIM}} provides a high-level overview of the rank-level architecture.

\filterfuselongreadacc performs three key tasks at the rank level: 
\begin{itemize}[leftmargin=*]
    
    \item Processing inputs from the host device in the correct format and sending them to the appropriate memory locations.

    \item Providing instructions, addresses, data, IDs, and masks to control the lower levels, stalls, and the results.

    \item Tracking read/reference pairings and implementing the edit-counting, summation, and comparison to the edit-distance threshold.

\end{itemize}

At the rank level, \filterfuselongreadacc uses multiple \counttcam{}s (\Gcircled{1} in \fig{\ref{fig:rank_architecture-filterFuse-proposed_architecture_hardware_accelerator_SSCIM}}) to collect the results of bank groups and calculate/count \#edits in a word set of the input read. A \counttcam detects the pattern and assigns the edit number of the detected patterns. Currently, \filterfuselongreadacc uses a 4-bit wide \tcam for each \counttcam{}. \filterfuselongreadacc can use this \counttcam to perform the final check for accepting or rejecting a read-reference pair in \geneguardianlongreadalg (line 22 in \alg{\ref{alg:LongGeneGuardian_Algorithm-FilterFuse_algorithm-proposal_and_architecture}}). This is also compatible with patterns in previous filters such as \shd, which required splitting the final bit-vector into segments of k=4 bits. Final masking on the output of \counttcam{}s (\Gcircled{2}) and a multi-operand adder (\Gcircled{3}) finalize the discovery of edits in the input read via \filterfuselongreadacc based on \geneguardianlongreadalg.

At the rank level, \filterfuselongreadacc divides a word over the bank groups such that each part goes to a different tile, as is the case in \dram~\cite{Kim2012ISCA_SALP_SubarraylevelparallelismDRAM} (\Gcircled{4}). This writing scheme ensures that different parts of the read/reference are written to different sub-arrays. In filtering algorithms, this means each processing element only has access to a small part of the read/reference. Therefore, algorithms such as \geneguardianlongreadalg that require larger segments to be examined by a single processing element require multiple words to be written before starting the algorithm. We call the number of required words words-per-bank (WPB) hereafter. To support a WPB of larger than 1, \filterfuselongreadacc implements a series of input buffers. These input buffers require the host device to provide read/reference sequences in the correct order without the need for excessive pre-processing. To fill the buffers, consecutive words belonging to a read-reference pairing are loaded into the input buffer sequentially. Once the buffer is filled with a single word set (i.e., 4 words), \filterfuselongreadacc empties them in parallel.

Since in a true \cim architecture, such as \filterfuselongreadacc, the first base pair of the read does not always coincide with the start of a word-set, \filterfuselongreadacc masks off the part that does not belong to the read-reference pairing. \filterfuselongreadacc receives this mask through the address bus and passes this mask to the sub-arrays alongside the read-sequence data (\Gcircled{5} in \fig{\ref{fig:rank_architecture-filterFuse-proposed_architecture_hardware_accelerator_SSCIM}}). \filterfuselongreadacc does not use the mask when filling the input buffers.

The rank level controller (\circled{6}) also handles the parallelism of new word sets via IDs, result-ready signals, and conservative buffering. For example, the rank-input-controller can load in a new word set during the computation of the previous word set as the sub-arrays can operate independently. Depending on the length of the read/reference pair, this can either be the next word set belonging to the same pairing or the first word of the next pairing.  The least significant bits (LSBs) of the ID of pairs are passed along with the input data to the sub-arrays, while the most significant bits (MSBs) are stored at the rank level in the ID MSB buffer, which is indexed by the LSB of the ID. This helps \filterfuselongreadacc to reduce the width of the ID busses. When the computation is completed, the results are returned, and the LSBs of the ID are matched up with the MSB again.

%% file: Sections/11_algorithm_mapping.tex
\subsection{Data Mapping in  \filterfuselongreadacc}
\label{subsec:sscim_algorithm_mapping_acchwsscim-proposed_architecture_hardware_accelerator_SSCIM}

\filterfuselongreadacc handles both references and read sequences inside the memory. For reference sequences, \filterfuselongreadacc assumes reference sequences are already stored in memory after being split up into word-sets. This is a reasonable, common assumption that can also be achieved easily as a pre-processing step if it is not the case.

\begin{figure}[htbp]
\centering
    \includegraphics[width=1\linewidth]{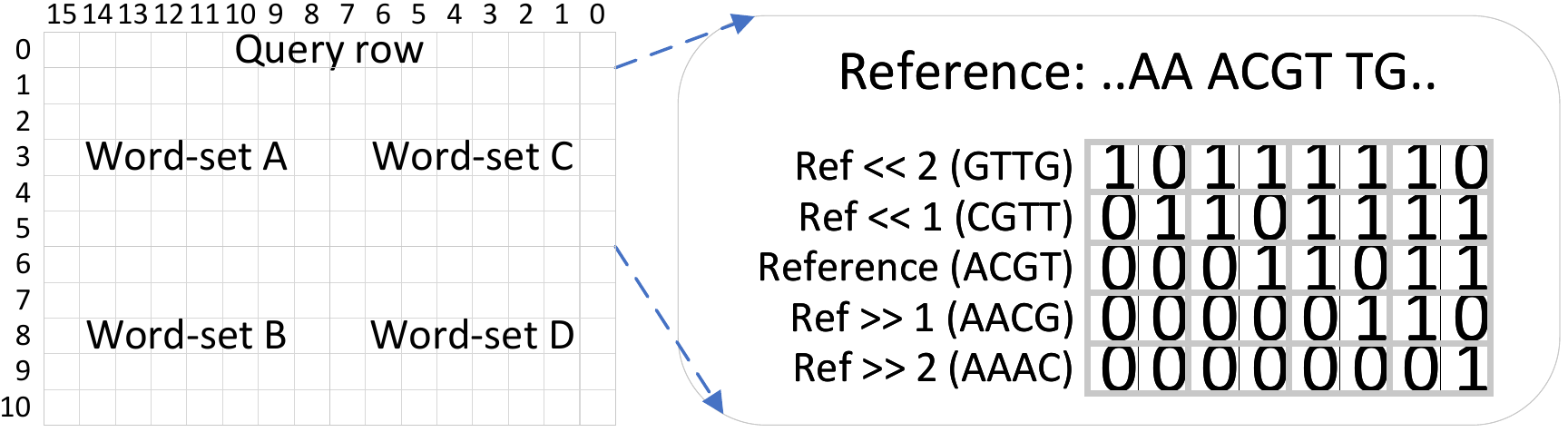}
    \caption{Reference mapping on a small crossbar.}
    
    \label{fig:reference_mapping-read_and_reference_mapping-sscim_algorithm_mapping_acchwsscim-proposed_architecture_hardware_accelerator_SSCIM-proposed_architecture_hardware_accelerator_SSCIM}
\end{figure}

Specifically, \filterfuselongreadacc assumes that each word-set is stored in a single row in the memory, with shifted versions of the word-set being in its adjacent crossbar rows within the same crossbar. \filterfuselongreadacc reserves the first row of each tile for the read sequence, hereafter referred to as the 'query row.' \fig{\ref{fig:reference_mapping-read_and_reference_mapping-sscim_algorithm_mapping_acchwsscim-proposed_architecture_hardware_accelerator_SSCIM-proposed_architecture_hardware_accelerator_SSCIM}} demonstrates this mapping, where we assume a $16 \times 16$ crossbar, containing parts of 6 word-sets for evaluating an edit distance of 2. The zoomed-in version of one word-set highlights how \filterfuselongreadacc stores the (shifted) references. Note that \fig{\ref{fig:reference_mapping-read_and_reference_mapping-sscim_algorithm_mapping_acchwsscim-proposed_architecture_hardware_accelerator_SSCIM-proposed_architecture_hardware_accelerator_SSCIM}} does not account for interleaving for the sake of clarity.

\begin{figure}[htbp]
\centering
    \includegraphics[width=1\linewidth]{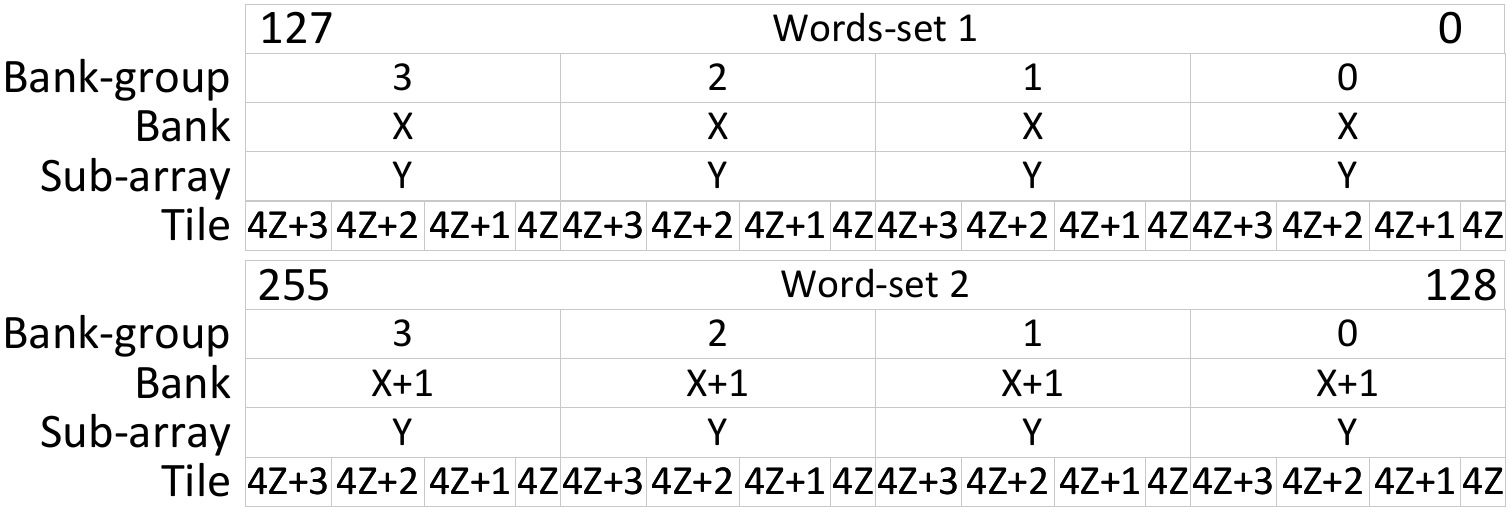}
    \caption{Writing scheme for two consecutive parts of the reference.}
    \label{fig:high_level_writing_wordsets_reference-read_and_reference_mapping-sscim_algorithm_mapping_acchwsscim-proposed_architecture_hardware_accelerator_SSCIM}
\end{figure}

For optimal performance, \filterfuselongreadacc also assumes that consecutive word-sets are mapped to different banks to avoid contention as much as possible. As discussed, reference sequence mapping is considered a pre-processing step; therefore, this assumption does not affect the overall performance of \filterfuselongreadacc. \fig{\ref{fig:high_level_writing_wordsets_reference-read_and_reference_mapping-sscim_algorithm_mapping_acchwsscim-proposed_architecture_hardware_accelerator_SSCIM}} presents a high-level overview of the sub-division of two consecutive word-sets for the reference. Here, we assume a hardware configuration with 32-bit words, 4 bank groups, tiles with 8 \sa{}s, and a 4-WPB writing scheme.

\filterfuselongreadacc only requires a simple 2-bit data encoding for pre-processing, a standard in genomics hardware accelerators~\cite{mansouri2022genstore, GenDP-ISCA2023-Das, cali2020genasm}. This pre-processing step is a one-time overhead in \filterfuselongreadacc for the preparation of reference sequences by the host device.

Since the read sequences come from the input buffer, \filterfuselongreadacc handles them directly and does not assume them being stored in the memory crossbar. Therefore, for each read sequence, \filterfuselongreadacc spreads the contents of the input buffer over the bank groups, writing the values to a set of tiles in a single sub-array per bank. We designed \filterfuselongreadacc's controller to efficiently handle internal data distribution for the read sequences. This approach negates the necessity for external input data acceleration, allowing the host device to oversee pre-alignment filtering with minimal overhead. Note that the host device does introduce some overhead in sequence preparation. Yet, this overhead remains minimal, aligning with other well-optimized \cim designs. Specifically, the host device mainly pinpoints \#$N_{segment}$ (using $T$ in \alg{\ref{alg:LongGeneGuardian_Algorithm-FilterFuse_algorithm-proposal_and_architecture}}), final Match count (Lines 20 to 22 in \alg{\ref{alg:LongGeneGuardian_Algorithm-FilterFuse_algorithm-proposal_and_architecture}}), and sets sequence start points (pointer to the start position of Read and Reference in Input of \alg{\ref{alg:LongGeneGuardian_Algorithm-FilterFuse_algorithm-proposal_and_architecture}}), a standard step in well-optimized \cim designs due to separate file sourcing. From there, \filterfuselongreadacc takes over, handling sequence segments and comparisons internally.

\begin{figure}[htbp]
\centering
    \includegraphics[width=1\linewidth]{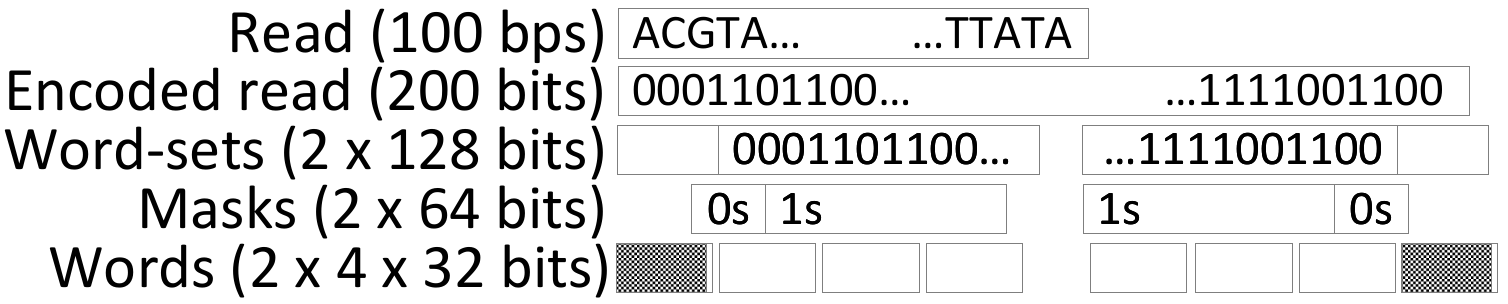}
    \caption{Reads subdivision into word-sets with masking.}
    \label{fig:Reads_to_words_or_wordsets-read_and_reference_mapping-sscim_algorithm_mapping_acchwsscim}
\end{figure}

\fig{\ref{fig:Reads_to_words_or_wordsets-read_and_reference_mapping-sscim_algorithm_mapping_acchwsscim}} illustrates an example of the subdivision of an input read sequence by \filterfuselongreadacc into words and word-sets alongside the masking process. Here, we assume a 100 \bp{}s read encoded in 200 bits.

\begin{figure*}[htbp]
\centering
    \includegraphics[width=1\linewidth]{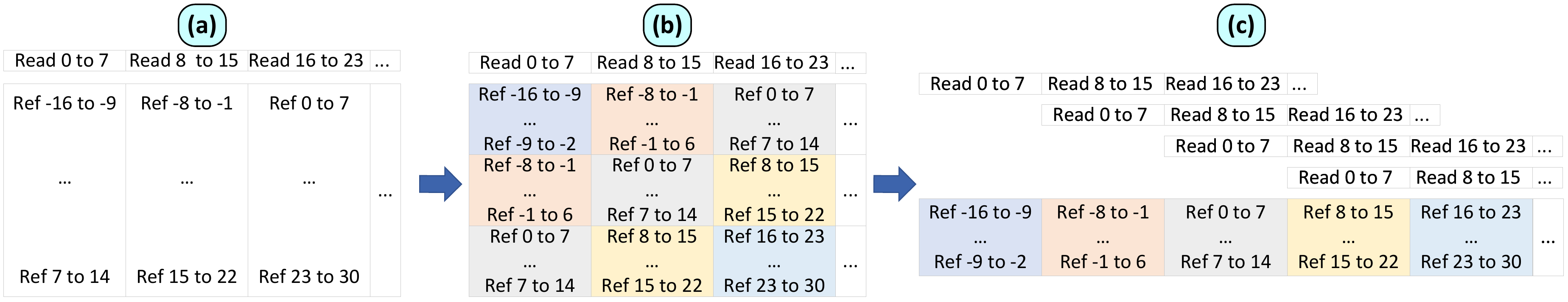}
    \caption{\filterfuselongreadacc's (a) high-level data mapping, (b) split data into shift-sets, and (c) the memory-optimized configuration.}
    \label{fig:example_long_read_shift_originalSplitAndFinal-long_read_compatibility_discussion-proposed_architecture_hardware_accelerator_SSCIM}
\end{figure*}

\filterfuselongreadacc divides the read over two word-sets of 128 bits. It then finds the address of the word-set containing the first bit of the read sequence using the mentioned fixed equations. Note that since the start of the read does not necessarily coincide with the start of a word-set, \filterfuselongreadacc masks the first part of the first and last part of the second word-set. \filterfuselongreadacc determines this mask by subtracting the offset of the first base pair of the word-set from the offset of the read (seeding location). We define the offset as the position of the reference segment with respect to the first base pair in the evaluated reference genome. The controller uses simple equations to find this offset. \filterfuselongreadacc then uses this local offset to find the length of the leading mask. \filterfuselongreadacc always writes the read to the query row of the tiles such that the columns line up with the addressed part of the reference sequence.

%% file: Sections/14_long_read_compatibility.tex
\subsection{Long Read Compatibility}
\label{subsec:long_read_compatibility_discussion-proposed_architecture_hardware_accelerator_SSCIM}

In previous sections, we assumed all shifted references are written to the same tile for ease of explanation. However, this assumption has two limitations for long reads. First, it limits the maximum edit distance \filterfuselongreadacc can support to the rows within each of its tiles. Since crossbars have limited dimensions (due to factors such as increasing read/write current requirements and leakage currents as the dimensions are scaled-up), the maximum supported edit distance becomes limited. Second, it demands an unreasonable memory capacity as the number of shifted references required for long reads exceeds 50 thousand shifts.

Therefore, to enable processing long reads, \filterfuselongreadacc splits up the evaluation of the different shifts into parts and processes them in different sub-arrays and over multiple iterations. \filterfuselongreadacc can then aggregate the results at the rank level before calculating the edits. This requires \filterfuselongreadacc to adopt two types of changes on top of the simplified examples of previous sections: (1) hardware changes at the sub-array and rank levels and (2) software changes at the input level.

Having the third observation in \geneguardianlongreadalg in mind (see \sect{\ref{sec:proposed_algorithm_SSCIM}}), \filterfuselongreadacc exploits the trade-off between the required memory capacity and the endurance of or the necessary write operations in the system, by splitting up read-sequences into segments of base pairs as we are splitting up the shifts. This capability separates \filterfuselongreadacc from the architecture for \rattlesnakejakesamos. \fig{\ref{fig:example_long_read_shift_originalSplitAndFinal-long_read_compatibility_discussion-proposed_architecture_hardware_accelerator_SSCIM}} presents an example of this trade-off, where the numbers indicate the ranges of bits that are evaluated in each shift set.

\fig{\ref{fig:example_long_read_shift_originalSplitAndFinal-long_read_compatibility_discussion-proposed_architecture_hardware_accelerator_SSCIM}}-(a)  represents a case where the segments of the read sequence are compared to the entire set of shifted references in the same tile. The read sequence is split up into segments, as well as in sections of 8 shifts, which are evaluated separately. We refer to these partitions as shift sets. We observe that shift sets that share a diagonal evaluate the same sections of the reference sequence. Therefore, in \fig{\ref{fig:example_long_read_shift_originalSplitAndFinal-long_read_compatibility_discussion-proposed_architecture_hardware_accelerator_SSCIM}}-(c), \filterfuselongreadacc only stores one of each shift-set, while the reads shifted with respect to the shift-sets.

The host device handles the shift of reads for \filterfuselongreadacc and prevents incorrect results from the shifted outputs of sub-array results by passing the shift value alongside the pairing ID. \filterfuselongreadacc saves the sub-array result for each shift-set in an \gateandop buffer, which is placed alongside the sum buffer (\Gcircled{7} in \fig{\ref{fig:rank_architecture-filterFuse-proposed_architecture_hardware_accelerator_SSCIM}}). This way, \filterfuselongreadacc accumulates the partial bit-vector results of all shift sets in this buffer by performing a bitwise \gateandop-operation between its stored value and the incoming result. When all shift sets of the read segment have been evaluated, \filterfuselongreadacc uses contents of the \gateandop-buffer as input to the \counttcam (\Gcircled{1} in \fig{\ref{fig:rank_architecture-filterFuse-proposed_architecture_hardware_accelerator_SSCIM}}). The rest of the procedure is identical to what was discussed previously. To prevent incorrect results from the shifted outputs of sub-array results (happening as the read sequence is shifted with respect to the start of each word set), we pass the shift value alongside the pairing ID.

%% file: Sections/13_differences_software_vs_hardware_Implementation.tex
\subsection{\geneguardianlongreadalg on Software vs. on \filterfuselongreadacc}
\label{subsec:difference_sw_hw-proposed_architecture_hardware_accelerator_SSCIM}

\begin{figure*}[htbp]
\centering
    \includegraphics[width=0.7\linewidth]{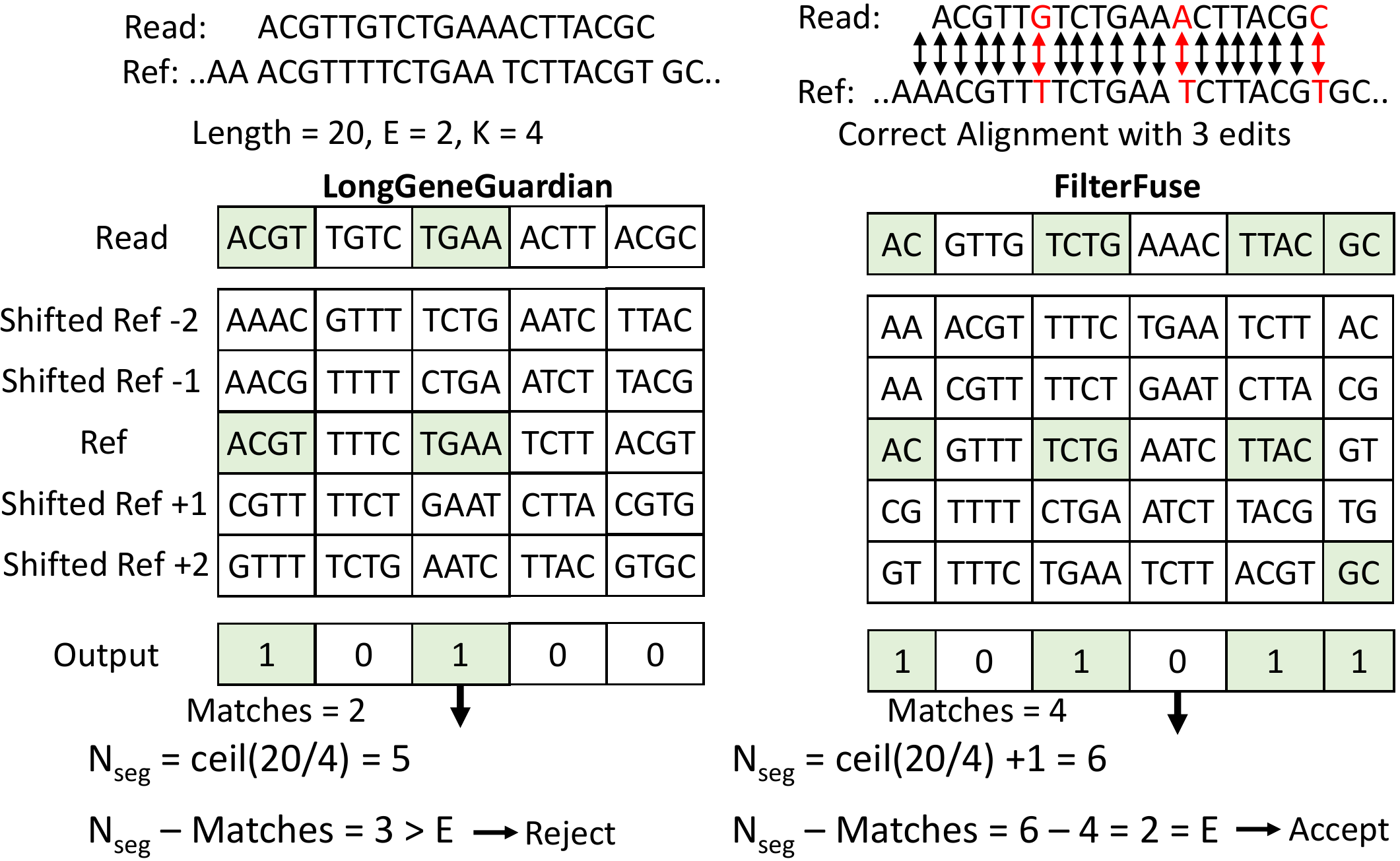}
    \caption{\geneguardianlongreadalg on Software vs. on \filterfuselongreadacc.}
    \label{fig:example_difference_sw_hw-difference_sw_hw-proposed_architecture_hardware_accelerator_SSCIM}
\end{figure*}

\geneguardianlongreadalg is a standalone pre-alignment filtering algorithm operating on both software, like \cpu{}s, and on \filterfuselongreadacc. However, the accuracy of \geneguardianlongreadalg might differ between the two. This is because, unlike on software, where segments always start at the start of the sequence, on \filterfuselongreadacc, sequences can begin mid-segment, as a true \cim accelerator where references have already stored in a fixed position in the memory elements. \fig{\ref{fig:example_difference_sw_hw-difference_sw_hw-proposed_architecture_hardware_accelerator_SSCIM}} demonstrates an example for this scenario.

Consequently, \geneguardianlongreadalg on software always contains the minimum number of segments for a given reads sequence, while the implementation on \filterfuselongreadacc can contain additional segments. On the one hand, this allows the \filterfuselongreadacc to detect more edits than its software counterpart. On the other hand, this increases the likelihood of random matches occurring since the number of evaluated base pairs remains the same, making the first/last segments shorter than the intended segment length. Our internal investigations on this difference show that on real datasets, \geneguardianlongreadalg on software is slightly more accurate than \filterfuselongreadacc. However, since \geneguardianlongreadalg on software still does not reach the accuracy or performance of \sneakysnake (our \sota baseline in \sect{\ref{sec:evaluations_and_results}}), we do not directly compare \geneguardianlongreadalg on software in the upcoming sections. 

%% file: Sections/16_evaluation_methodology_setup.tex
\section{Evaluation Methodology} 
\label{sec:experimental_setup_and_evaluation_methodology}

\noindent\textbf{Implementations \& Models.} 
We implement \geneguardianlongreadalg on software (\cpp) for its accuracy evaluations. We use a cycle-accurate RTL-based implementation to verify the functionality of \filterfuselongreadacc. The analog components (i.e., \reram-based crossbars and \tcam{}s) are memory models from actual \reram crossbars in TSMC \SI{40}{\nano\meter} CMOS technology~\cite{kim2016multistate-hfo2tiox}, from the EU project MNEMOSENE~\cite{MNEMOSENE-TUDelftEU}. The \dram-based crossbars are from \simdram~\cite{hajinazar2021simdram, seshadri2017ambit}. The additional components are also designed using TSMC \SI{40}{\nano\meter} technology node in Synopsis Design Compiler~\cite{synopsys}. We run all our experiments on a 12-core server with \SI{16}{\giga\byte} memory, \teslagpu \gpu{}s, and a \intelxeonqce operating at \SI{2.4}{\giga\hertz}.

\noindent\textbf{Baselines.} We use \edlib~\cite{vsovsic2017edlib} on \cpu (\alignercpualignment) for the golden standard alignment results for accuracy evaluations. For end-to-end evaluations, we feed the output of each filter to \edlib. We mainly compare \filterfuselongreadacc with open-sourced \sneakysnake (SS)~\cite{SneakySnake} on \cpu (\sscpufilter), as the only \sota pre-alignment filter for long-reads\footnote{Neither the \fpga implementation nor the \gpu one supports long reads.}. We used two versions of \filterfuselongreadacc: (1) using \cmos (\filterfusecmos), where the tiles are \sota \dram-based and \tcam{}s are \sota \sram-based ones, and (2) using \reram (\filterfusereram), where both tiles and \tcam{}s are \reram-based ones. Previous well-optimized \cim accelerators only support short-read pre-alignment filtering. However, to have a crude estimation of their performance potential for long reads if they could support them, we consider two \cim filters for short reads (\grimthreedstacked~\cite{kim2018grim} and \rattlesnakejakereram~\cite{shahroodi2023RattlesnakeJake} on 3D-stacked- and \reram-based memories) in our comparison for filtering performance (\sect{\ref{subsec:filtering_throughput_and_execution_time_or_speed-evaluations_and_results}}). We cut the long sequences into smaller, non-overlapping chunks for these comparisons, later merging chunk results for a holistic long-read analysis. This is similar to our analysis of data movement overhead for \gpusneakysnake in \sect{\ref{subsec:Limitations_sota_pre_alignment_filters_long_reads-motivation}}.

\begin{figure*}[htbp]
\centering
    \includegraphics[width=0.8\linewidth]{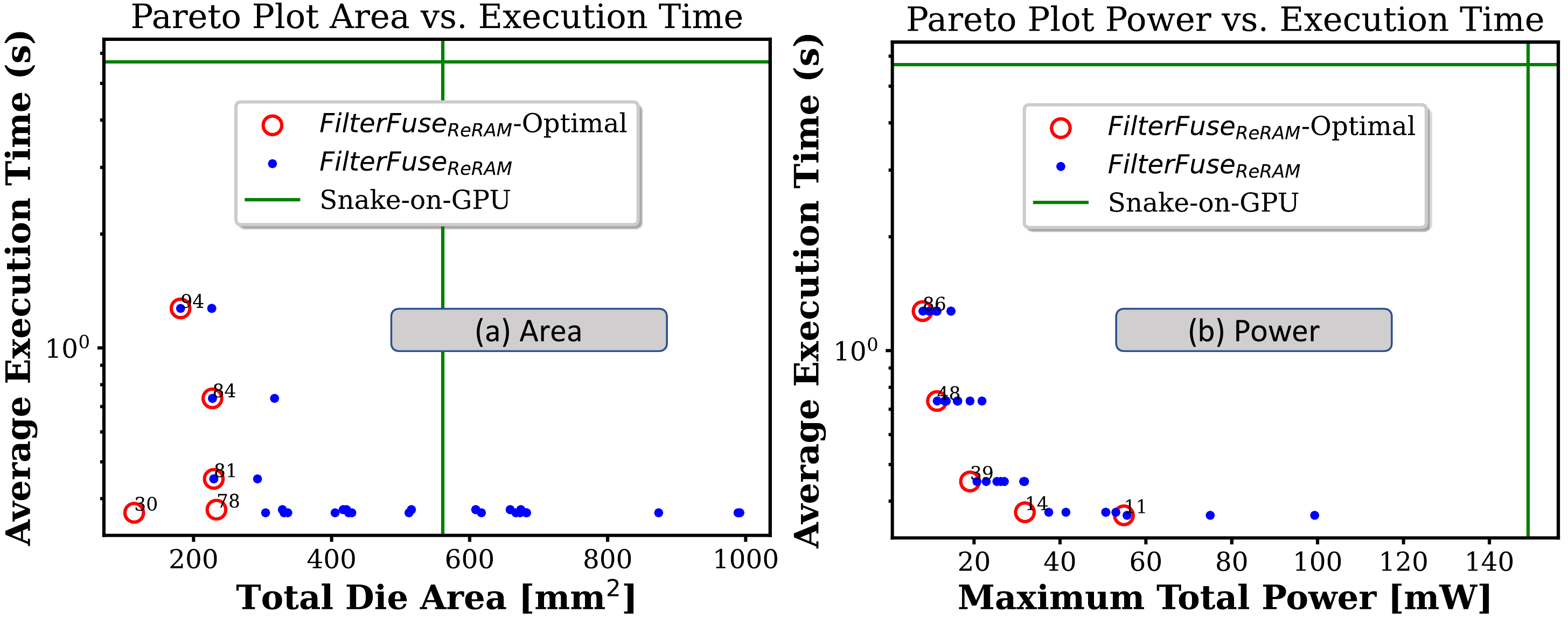}
    \caption{Pareto plots for performance per area and power.}
    
    \label{fig:pareto_perf_area_and_power-design_space_exploration-evaluations_and_results}
\end{figure*}

\begin{figure*}[htbp] 
\centering
    \includegraphics[width=0.8\linewidth]{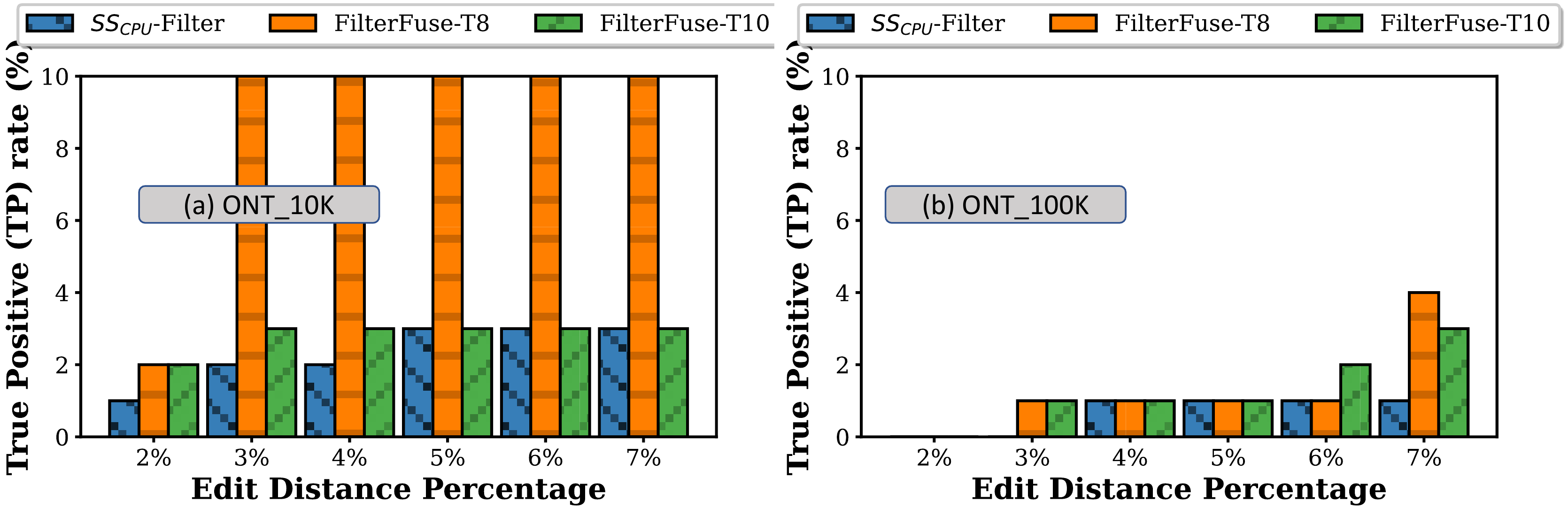}
    \caption{\tp rate of filtering with different segment lengths.} 
    \label{fig:TP_rate_TruePositive_longReads_filtering_PacBio9810KAndPacBio98100K_SneakySnakeCPU_vs_FilterFuse8_vs_FilterFuse10-accuracy_filtering_accuracy_analysis-evaluations_and_results}
\end{figure*}

\noindent\textbf{Datasets.} We use \pbsimthree~\cite{PBSIM3-longReads_simulator-Paper, PBSIM3-longReads_simulator-Code} to produce our main two datasets of long read-reference pairs from real datasets. We leaned on the simulator for our datasets, as it is both accurate and a standard approach in prior studies~\cite{alser2022molecules-FrommoleculestogenomicvariationsAcceleratinggenomeanalysis, SneakySnake}. While we have results from real datasets as well, controlling all facets of real datasets downloaded from the NCBI database is challenging. Thus, we consider the simulator as the optimal way to measure \filterfuselongreadacc’s potential while being dataset-dependent. Our results in this draft only focus on Ultra-Long Reads from \ont (as defined in \cite{alser2022molecules-FrommoleculestogenomicvariationsAcceleratinggenomeanalysis}) with +98\% sequencing accuracy. This accuracy justifies our considered \roi (See \sect{\ref{sec:background_and_relatedWork}}) in the upcoming results. We also evaluated datasets from \pacbio \clr~\cite{alser2022molecules-FrommoleculestogenomicvariationsAcceleratinggenomeanalysis} with 80\% accuracy that changes the \roi to $\sim$25\%. However, we did not present our findings for \pacbio \clr as they show an increase in the overall end-to-end execution time of alignment as defined in \sect{\ref{sec:background_and_relatedWork}}. This is because the filtering overhead is not justified for less accurate reads. This aligns with those in prior works~\cite{SneakySnake, kim2018grim}. Fortunately, there is a large momentum towards accurate long-read sequencing~\cite{alser2022molecules-FrommoleculestogenomicvariationsAcceleratinggenomeanalysis, cali2020genasm, senol2019nanopore-gagan16}. We then feed these reads to \minimap~\cite{li2018minimap2}, a \sota read mapper supporting long reads. We use the output to retrieve seed locations, by which we can retrieve the reference sequences corresponding to the read with SAM-tools~\cite{Samtools2021}. We call these datasets \onttenk and \onthunk, with reads of length 10\kbp and 100\kbp, respectively.

%% file: Sections/17_evaluations_and_results.tex
\section{Evaluation Results} 
\label{sec:evaluations_and_results}

\subsection{Design Space Exploration} \label{subsec:design_space_exploration-evaluations_and_results}

\fig{\ref{fig:pareto_perf_area_and_power-design_space_exploration-evaluations_and_results}}-(a) and -(b) present the Pareto optimal design space exploration of \filterfuselongreadacc for average execution time against total die area of \filterfuselongreadacc and its maximum total power, respectively. We mark most attractive designs that strike a sweet spot in the tradeoff between execution time and the corresponding metric with red circles in \fig{\ref{fig:pareto_perf_area_and_power-design_space_exploration-evaluations_and_results}} and call them area-optimized and power-optimized designs. Numbers label the Pareto optimal configurations, each of which is a different hardware configuration we tested for \filterfuselongreadacc, but the full list is not presented for better readability.

From \fig{\ref{fig:pareto_perf_area_and_power-design_space_exploration-evaluations_and_results}}-(a) we make two observations. First, all of the area-optimized configurations have a smaller area than that of our \gpu. Second, configuration \#81 strikes a great balance of power while still optimizing for the area. More investigations also reveal that for the area-optimized designs, the configurations with large tile dimensions are favored. This is expected due to their smaller tile and control logic area.

\begin{figure*}[htbp] 
\centering
    \includegraphics[width=0.8\linewidth]{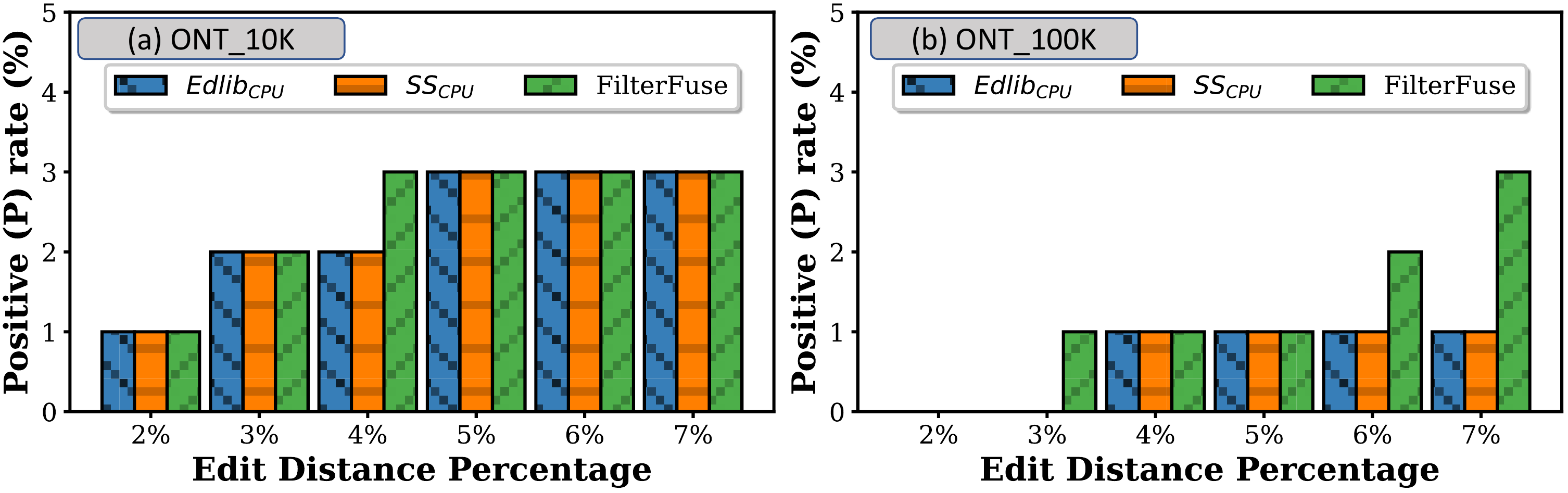}
    \caption{Positive rate of filtering with the best segment.} 
    \label{fig:P_rate_Positive_longReads_filtering_PacBio9810KAndPacBio98100K_SneakySnakeCPU_vs_FilterFuse8_vs_FilterFuse10-accuracy_filtering_accuracy_analysis-evaluations_and_results}
\end{figure*}

\begin{figure*}[htbp] 
\centering
    \includegraphics[width=0.8\linewidth]{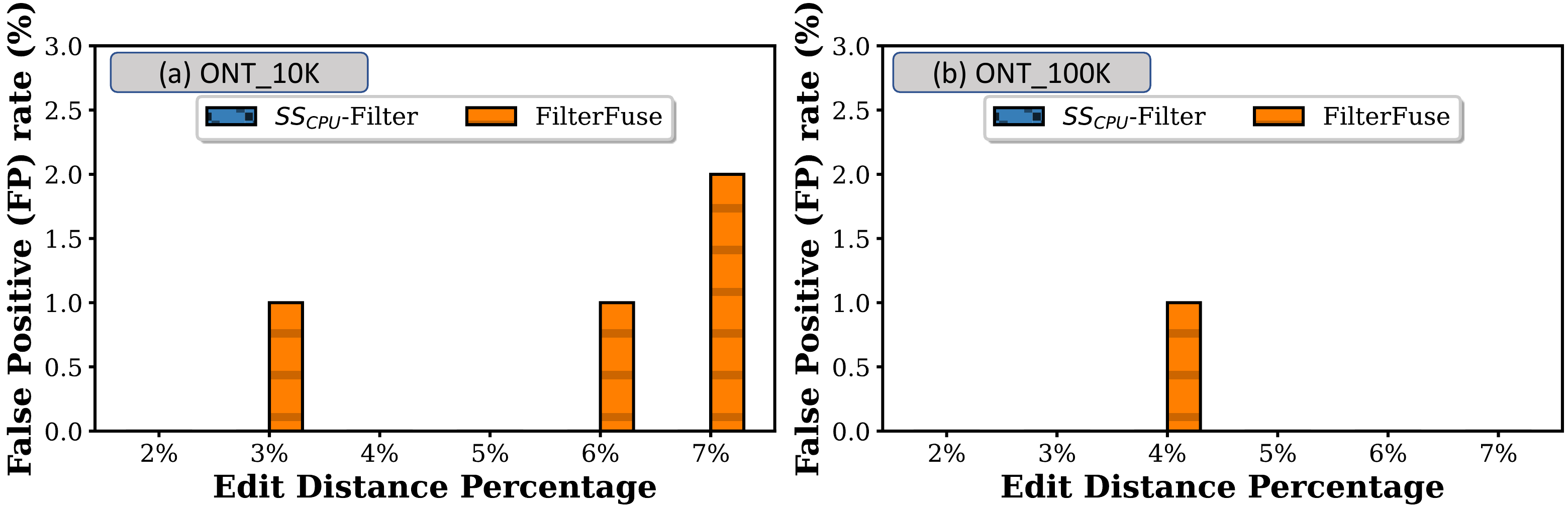}
    \caption{\fp rate of long reads filtering with the best segment.}
    \label{fig:FP_rate_FalsePositive_longReads_filtering_PacBio9810KAndPacBio98100K_SneakySnakeCPU_vs_FilterFuse8_vs_FilterFuse10-accuracy_filtering_accuracy_analysis-evaluations_and_results}
\end{figure*}

From \fig{\ref{fig:pareto_perf_area_and_power-design_space_exploration-evaluations_and_results}}-(b) we make two observations. First, the power consumption is much lower for power-optimized configurations compared to that of the baseline \gpu. Second, configuration \#39 provides a prominent balance between area and power for the power-optimized Pareto optimal configurations. From further investigations, we also find that this optimization criterion favors configurations with large numbers of tiles per sub-array. We expect this as these configurations have fewer active tiles at a given moment, as each sub-array only uses part of its available tiles at a time.

\subsection{Filtering Accuracy} \label{subsec:accuracy_filtering_accuracy_analysis-evaluations_and_results}

\fig{\ref{fig:TP_rate_TruePositive_longReads_filtering_PacBio9810KAndPacBio98100K_SneakySnakeCPU_vs_FilterFuse8_vs_FilterFuse10-accuracy_filtering_accuracy_analysis-evaluations_and_results}} comapres the \tp rate of \sneakysnake with \filterfuselongreadacc using different segment lengths for our datasets.

We observe that the best accuracy varies with the edit distance, with a segment size of 8 \bp{}s being close to the optimal rate in our \roi. For this reason, in the upcoming sections, we will only present the results of this segment length.

\fig{\ref{fig:P_rate_Positive_longReads_filtering_PacBio9810KAndPacBio98100K_SneakySnakeCPU_vs_FilterFuse8_vs_FilterFuse10-accuracy_filtering_accuracy_analysis-evaluations_and_results}} presents the positive rate (P-rate = \fp-rate + \tp-rate) of optimal \filterfuselongreadacc, i.e., an indication of evaluated pairs that require alignment.

We observe that \filterfuselongreadacc achieves the same or a maximum of 1\% higher P-rate compared to \sneakysnake and \edlib.

\fig{\ref{fig:FP_rate_FalsePositive_longReads_filtering_PacBio9810KAndPacBio98100K_SneakySnakeCPU_vs_FilterFuse8_vs_FilterFuse10-accuracy_filtering_accuracy_analysis-evaluations_and_results}} presents the \fp rate of \filterfuselongreadacc and \sneakysnake for various edit distances and datasets.

We observe that \filterfuselongreadacc retains a low \fp rate of $<2\%$ compared to  \sneakysnake, which achieves the lowest \fp rate among all previous filters.

From these results, we conclude that  \filterfuselongreadacc is an effective and accurate filter for long reads, achieving an accuracy as close as the \sota pre-alignment filter, \sneakysnake. We note that we observed that using non-overlap chunks of short reads for long-read filtering reduces accuracy, as confirmed by prior studies~\cite{xin2015shifted-SHD, alser2019shouji, SneakySnake}. Hence, we didn't showcase the accuracy of our short-read \cim accelerators.

\subsection{Filtering Speed} \label{subsec:filtering_throughput_and_execution_time_or_speed-evaluations_and_results}

\begin{figure*}[htbp] 
\centering
    \includegraphics[width=0.8\linewidth]{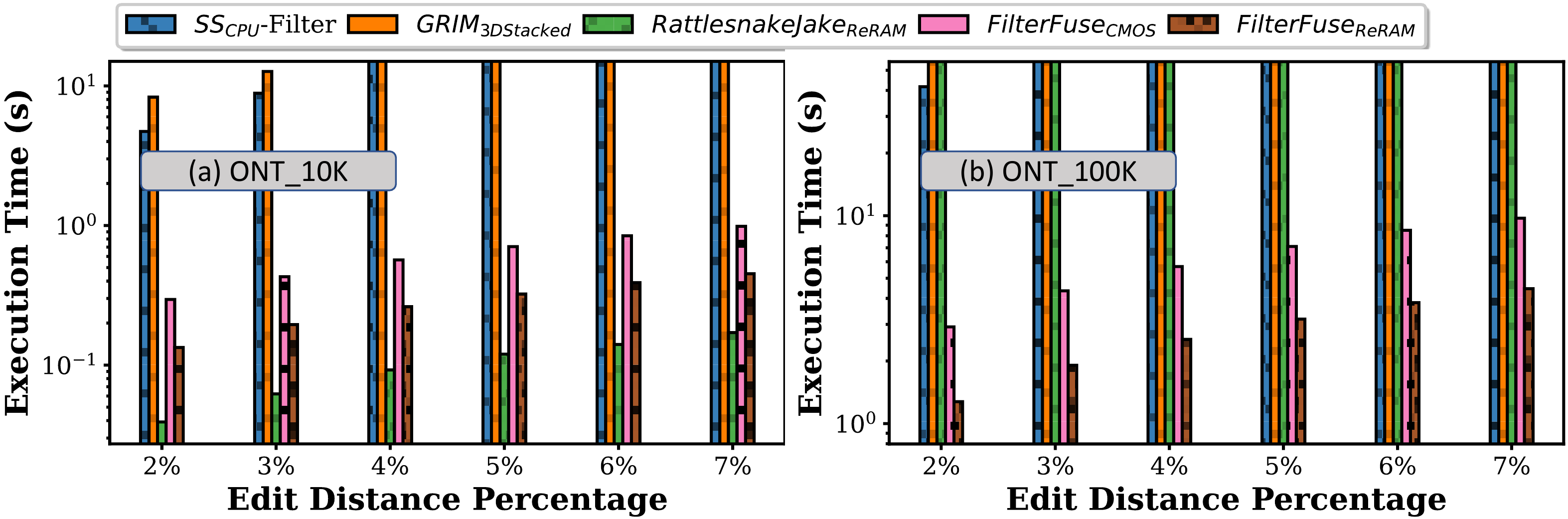}
    \caption{Speed of long reads filtering.}
    \label{fig:partial_filtering_ExecutionTime_longRead_PacBio9810KAndPacBio98100K_SneakySnakeCPU_vs_FilterFuse8-filtering_throughput_and_execution_time_or_speed-evaluations_and_results}
\end{figure*}

\fig{\ref{fig:partial_filtering_ExecutionTime_longRead_PacBio9810KAndPacBio98100K_SneakySnakeCPU_vs_FilterFuse8-filtering_throughput_and_execution_time_or_speed-evaluations_and_results}} presents the execution time of \filterfuselongreadacc and our baselines over different edit thresholds and datasets.

\begin{figure*}[htbp] 
\centering
    \includegraphics[width=0.8\linewidth]{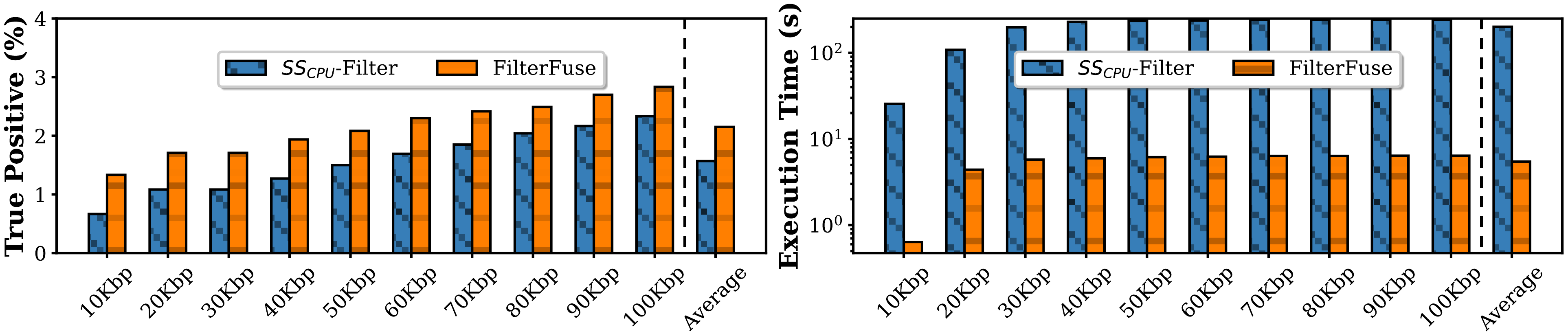}
    \caption{Sensitivity analysis for \tp rate and speed of long reads filtering.}
    \label{fig:sensitivityAnalysis_TP_And_FilteringSpeed_SneakySnakeCPU_FilterFuse-sensitivityAnalysis_TP_plus_FilteringThroughputorSpeed-evaluations_and_results}
\end{figure*}

\begin{figure*}[htbp]
\centering
    \includegraphics[width=0.8\linewidth]{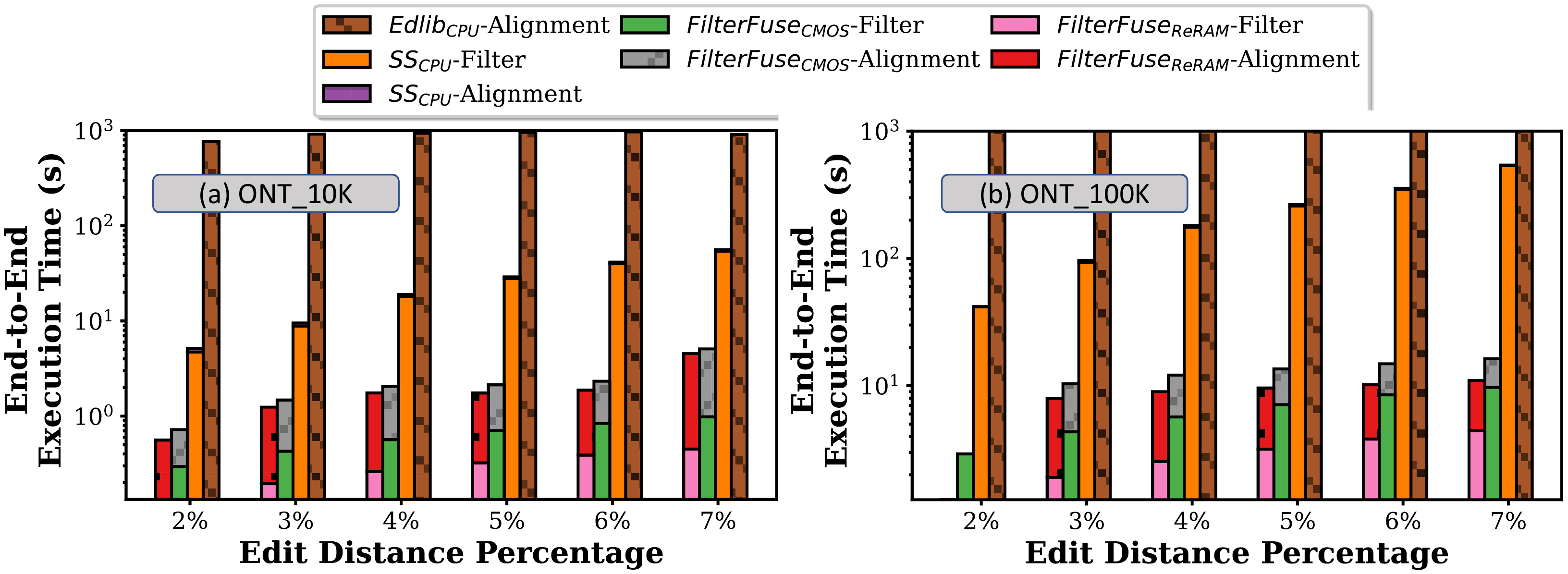}
    \caption{End-to-end speed of long reads filtering and alignment.}
    
    \label{fig:partial_EndtoEnd_ExecutionTime_longRead_PacBio9810KAndPacBio98100K_SneakySnakeCPU_vs_FilterFuse8-alignmentEndtoend_throughput_and_execution_time_or_speed-evaluations_and_results}
\end{figure*}

We make four main observations. First, \filterfuselongreadacc always outperforms \sscpufilter irrespective of the read length, edit threshold, and underlying technology. This improvement can go up to \timeImprovementFilteringGeneguardianlongreadalgOverSneakysnakeCPUUpto. Second, \filterfusereram outperforms \filterfusecmos. Further investigations show that this is because of the copy overhead used for \gatexorop in the underlying tile of \dram, even though the \sram-based \tcam{}s offset some of the performance overhead. Third, \filterfuselongreadacc consistently outperforms \grimthreedstacked irrespective of the read length, edit threshold, and underlying technology. The improvements over \grimthreedstacked are more significant than that of over \sscpufilter mainly due to the superior accuracy of \geneguardianlongreadalg over the underlying algorithm in \grimthreedstacked. Fourth, while \rattlesnakejakereram exhibits better performance with shorter long-reads like \onttenk, its efficacy diminishes with longer long-reads such as \onthunk due to architectural bottlenecks. The marginal advantage of \rattlesnakejakereram over \filterfuselongreadacc for shorter long-reads does not compensate for its lower accuracy and increased storage demands, as detailed in sections \sect{\ref{subsec:accuracy_filtering_accuracy_analysis-evaluations_and_results}} and \sect{\ref{subsec:long_read_compatibility_discussion-proposed_architecture_hardware_accelerator_SSCIM}}. We conclude that \filterfuselongreadacc is much faster than \sscpufilter mainly due to exploiting \cim to eliminate the data movement overhead and its lightweight algorithm. \filterfuselongreadacc also stands out among other \cim accelerators, especially considering the additional overheads of handling comparable data volumes.

\subsection{Sensitivity Analysis} \label{subsec:sensitivityAnalysis_TP_plus_FilteringThroughputorSpeed-evaluations_and_results}

We observe a clear dataset dependency in the results, mirroring patterns observed in other pre-alignment filters like \sneakysnake and \rattlesnakejakesamos. Our chosen datasets, distinguished by their unique lengths and error rates, were specifically selected for sensitivity analysis. However, we present a succinct sensitivity analysis, constrained by space, for \tp rate and filtering speed, mainly focusing on long reads that maintain high accuracy but differ in length. We leave a comprehensive sensitivity analysis for our forthcoming extended report. \fig{\ref{fig:sensitivityAnalysis_TP_And_FilteringSpeed_SneakySnakeCPU_FilterFuse-sensitivityAnalysis_TP_plus_FilteringThroughputorSpeed-evaluations_and_results}}  presents the outcomes of this analysis, both for the \tp rate (on the left) and filtering speed (on the right). We make two main observations.

First, \filterfuselongreadacc consistently achieves high accuracy, closely mirroring the \tp rate of \sscpufilter, as in \sect{\ref{subsec:accuracy_filtering_accuracy_analysis-evaluations_and_results}}. This indicates the robustness of \filterfuselongreadacc, regardless of long-read length. Second, \filterfuselongreadacc consistently surpasses the performance of \sscpufilter across all evaluated long-read lengths, underscoring its superior efficiency. We conclude that not only \filterfuselongreadacc is advantageous over \sota \sscpufilter and can be considered an accurate and fast pre-alignment filter for long-reads.

\begin{figure*}[htbp] 
\centering
    \includegraphics[width=0.8\linewidth]{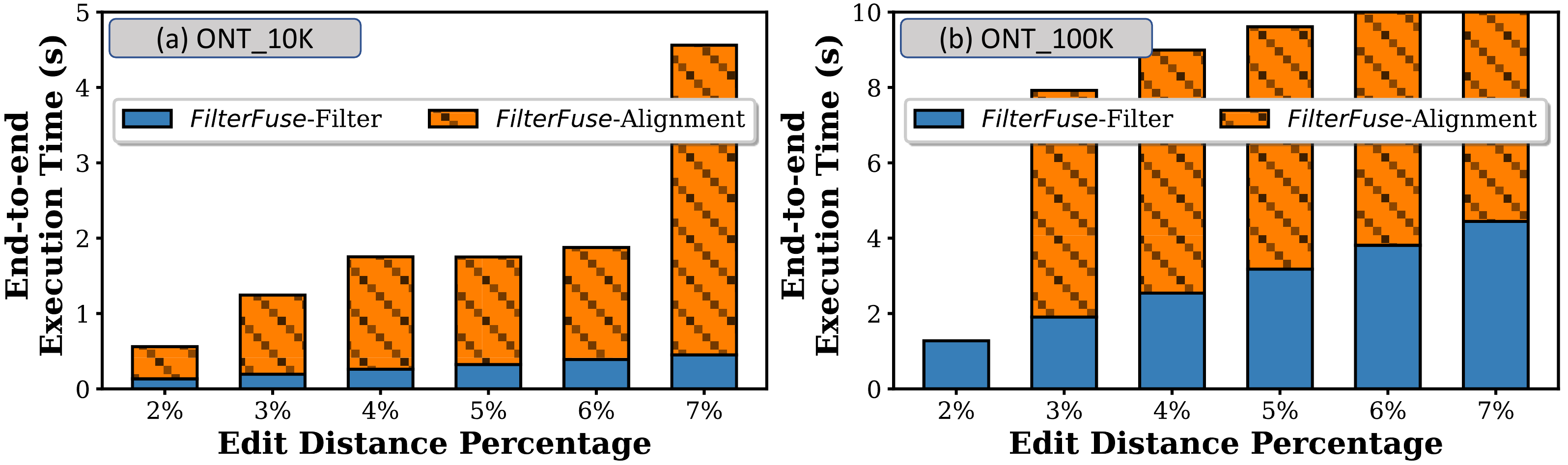}
    \caption{Filtering and alignment contribution in end-to-end execution time when using \filterfuselongreadacc .}
    \label{fig:partial_EndtoEnd_ExecutionTime_longRead_PacBio9810KAndPacBio98100K_FilterFuse_FilterContribution-alignmentEndtoend_throughput_and_execution_time_or_speed-evaluations_and_results}
\end{figure*}

\subsection{End-to-end Alignment Speed} \label{subsec:alignmentEndtoend_throughput_and_execution_time_or_speed-evaluations_and_results}

A filter that is faster than the other but has lower accuracy (higher \fp rate) might end up with higher end-to-end execution time when one also considers the time required for the alignment of the pairs that pass each filter. Therefore, it is necessary to compare the end-to-end execution, i.e., filtering and alignment for a given dataset, when comparing the effectiveness of a filter. \fig{\ref{fig:partial_EndtoEnd_ExecutionTime_longRead_PacBio9810KAndPacBio98100K_SneakySnakeCPU_vs_FilterFuse8-alignmentEndtoend_throughput_and_execution_time_or_speed-evaluations_and_results}} presents the end-to-end execution time for a filter followed by alignment over several edit thresholds and datasets. To better capture the trends (and relative execution time), we limit the y-axis to 1000s. The y-axis uses a logarithmic scale. Note that we omitted \grimthreedstacked and \rattlesnakejakereram as in \sect{\ref{subsec:filtering_throughput_and_execution_time_or_speed-evaluations_and_results}} because they do not support long reads (see \sect{\ref{sec:experimental_setup_and_evaluation_methodology}}). This makes a direct time comparison unfair and not present here.

We make two observations. First, for all datasets, edit-distance thresholds, and memory technologies, \filterfuselongreadacc shows an improvement of end-to-end execution time over \sneakysnake in the \roi. This improvement goes up to \timeImprovementEndtoendFilterfuselongreadaccOverSneakysnakeCPUUpto over \sneakysnake. Second, as expected, \filterfusereram improves the end-to-end time further than \filterfusecmos, while the difference is less than \timeImprovementEndtoendFilterfuseReRAMlongreadaccOverFilterfuseCMOSUpto. We conclude that even with the decrease in the accuracy (\sect{\ref{subsec:accuracy_filtering_accuracy_analysis-evaluations_and_results}}), \filterfuselongreadacc improves the performance of alignment significantly.

To determine whether \filterfuselongreadacc resolves the filtering bottleneck, we also examine the new distribution of filtering time and alignment time.  
\fig{\ref{fig:partial_EndtoEnd_ExecutionTime_longRead_PacBio9810KAndPacBio98100K_FilterFuse_FilterContribution-alignmentEndtoend_throughput_and_execution_time_or_speed-evaluations_and_results}}
presents this relative distribution over various edit distance thresholds and datasets.

We observe that alignment constitutes a minimum of  \alignmentContributionEndtoendFilterfuseLongreadMinimum of the end-to-end execution time, going as high as \alignmentContributionEndtoendFilterfuseLongreadUpto of the end-to-end execution time. This means that \filterfuselongreadacc improves the filtering step enough to move the bottleneck back to the long read alignment step, making the alignment the next computational step to focus on again.

\subsection{Area and Power Analysis} \label{subsec:area_and_power_analysis-evaluations_and_results}

\tab{\ref{tab:power_and_area_breakdown_ optimal_confi-area_and_power_analysis-evaluations_and_results}} presents the chip area and power consumption breakdown of the optimal configurations of \filterfuselongreadacc.

\begin{table}[htbp] 
\small
\scriptsize
\centering
\renewcommand{\tabcolsep}{1pt}
\begin{tabular}{ccc}
\hline
\multicolumn{1}{c|}{\textbf{Logic Unit}}  & \multicolumn{1}{c|}{\textbf{Area[$\mu m^2$]}} & \multicolumn{1}{c}{\textbf{Power[$mW$]}} \\ \hline\hline
Crossbars                                  & 130582105.29                                 & 3.93                                         \\
TCAMs                                      & 615813.12                                    & 0.00986                                      \\
Control Logics                             & 98032937.56                                  & 27.57                                        \\ \hline \hline
\multicolumn{1}{c|}{Total for FilterFuse} & \multicolumn{1}{c|}{229230855.97}            & \multicolumn{1}{c}{31.51}                   \\ \hline
\end{tabular}
\caption{Area and power breakdown of \filterfuselongreadacc.}
\label{tab:power_and_area_breakdown_ optimal_confi-area_and_power_analysis-evaluations_and_results}
\end{table}

We make three main observations. First, most chip area belongs to the crossbars and the tile-level control logic. The contributions of higher-level components are negligible, considering the total chip area. Second, the biggest contributor to the power is the tile-level control logic. Three, \tcam{}s add insignificant area and power consumption overheads.

\tab{\ref{tab:power_and_area_GPUSneakySnake_vs_proposal_optimal_confi_longreads-area_and_power_analysis-evaluations_and_results}} compare the chip area and power consumption of \filterfuselongreadacc with \sota design on our \gpu, where the maximum \gpu power is measured using nvidia-smi while running \gpusneakysnake. Note that area and power estimations of \filterfuselongreadacc components that were scaled down to the evaluation technology node are scaled pessimistically, leading to conservative estimates.

\begin{table}[htbp] 
\small
\scriptsize
\centering
\renewcommand{\tabcolsep}{1pt}
\begin{tabular}{|c|c|c|}
\hline
\textbf{Hardware} & \textbf{Area[$mm^2$]} & \textbf{Power[$W$]} \\ \hline
NVIDIA Tesla K80  & 561              & 149             \\ \hline
FilterFuse        & 229              & 31.5            \\ \hline
\end{tabular}
\caption{Area and power of \filterfuselongreadacc vs. \gpusneakysnake.}
\label{tab:power_and_area_GPUSneakySnake_vs_proposal_optimal_confi_longreads-area_and_power_analysis-evaluations_and_results}
\end{table}

We make two observations. First, \filterfuselongreadacc has a smaller overall chip area than our \teslagpu \gpu{}s. Second, \filterfuselongreadacc shows a lower maximum power consumption than \gpusneakysnake, i.e., a reduction of \powerReductionFilterfuseLongreadVsGpusneakysnake.

We conclude that \filterfuselongreadacc also has area and power advantages over a typical \gpu implementation.

%% file: Sections/19_relatedwork.tex
\section{Related Works} 
\label{sec:relatedWork}

We have already compared \filterfuselongreadacc extensively with \sota filter for long reads in \sect{\ref{sec:evaluations_and_results}}. This section briefly discusses the previous works on sequence alignment and (\cim) architectures for other genomics tasks.

\subsection{Sequence Alignment Acceleration}
\label{subsec:sequence_alignment_acceleration-relatedWork}

We have discussed the two directions previous works take to improve the sequencing alignment step directly in \sect{\ref{sec:background_and_relatedWork}}~\cite{lassmann2005kalign, crochemore2003subquadratic, fei2018fpgasw, luo2013soap3, notredame1996saga, crochemore2003subquadratic, daily2016parasail, georganas2015meraligner, banerjee2018asap, fei2018fpgasw, waidyasooriya2014fpga, chen2014accelerating, nishimura2017accelerating, liu2015gswabe, liu2013cudasw++, luo2013soap3, houtgast2015fpga, arram2013hardware}. \filterfuselongreadacc is orthogonal to all of these works as it is a step before the alignment and can be used with any sequence aligner. However, finding the best combination or the interface between \filterfuselongreadacc and a sequence aligner in this group is left for future work.

\subsection{\asic and \cim Accelerators for Genomics}
\label{subsec:asic_cim_based_genomics_accelerators-relatedWork}

Previous works~\cite{cali2020genasm, turakhia2018darwin, lou2020helix-gagan40, shahroodi2022krakenonmem, shahroodi2022demeter, mao2022genpip} propose various \asic and \cim architectures for different kernels in genomics pipelines.  \filterfuselongreadacc belongs to the works in this group. \filterfuselongreadacc accelerates pre-alignment filtering differentiating it from other works. In theory, \filterfuselongreadacc can be orthogonally used in conjunction with the works that require alignment. However, investigating the potential benefit of such a combination is left for future work.

%% file: Sections/20_conclusion.tex
\section{Conclusion} 
\label{sec:conclusion}

This paper proposes the first \cim architecture for pre-alignment filters of long reads, a major performance bottleneck in today's genome analysis of long reads. We call this \filterfuselongreadacc. \filterfuselongreadacc operates on a hardware-friendly algorithm, \geneguardianlongreadalg, that is also compatible with the requirements of a true \cim architecture: simple operations and no assumption on the data placement. Considering the larger genomics pipeline and industrial move towards long-read sequencing, \filterfuselongreadacc takes a large step in accelerating long-read genome analysis.